\documentclass[twocolumn, notitlepage, 10pt, aps, floatfix, showpacs, prb, citeautoscript, superscriptaddress]{revtex4-2}
\usepackage{amsfonts}
\usepackage{graphicx} 
\usepackage{float}
\usepackage{dcolumn}
\usepackage{bm}
\usepackage{color}
\usepackage{amsmath, amssymb}
\usepackage{bm}
\usepackage{xcolor}
\usepackage[colorlinks, citecolor={blue!50!black}, urlcolor={blue!50!black}, linkcolor={red!50!black}]{hyperref}
\usepackage{bookmark}
\usepackage{mhchem}
\usepackage[utf8]{inputenc}
 \usepackage[normalem]{ulem}

\usepackage{microtype}
\newcommand{\etilde}{{\widetilde \varepsilon\,}}
\newcommand{\Ptilde}{{\widetilde P\,}}
\newcommand{\eff}{\text{loc}}

\newcommand{\dtau}{ {\rm{d}\tau }}

\newcommand{\be}{\begin{equation}}
\newcommand{\ee}{\end{equation}}

\definecolor{darkgreen}{rgb}{0,0.5,0}
\definecolor{orange}{rgb}{1,0.5,0}
\definecolor{grey}{rgb}{.6,.6,.6}

\begin{document}

\title{Quantum Coulomb glass on the Bethe lattice }

\author{Izabella Lovas}
\affiliation{Department of Physics and Institute for Advanced Study, Technical University of Munich, D-85748 Garching, Germany}
\affiliation{Munich Center for Quantum Science and Technology (MCQST), Schellingstr. 4, D-80799 M\" unchen}
\author{Annam\'aria Kiss}
\affiliation{Institute for Solid State Physics and Optics, Wigner Research Centre for Physics, POBox 49, H-1525 Budapest, Hungary}
\author{C\u{a}t\u{a}lin Pa\c{s}cu Moca}
\affiliation{Department of Theoretical Physics, Institute of Physics, Budapest University of Technology and Economics, M\H uegyetem rkp. 3., H-1111 Budapest, Hungary}
\affiliation{Department  of  Physics,  University  of  Oradea,  410087,  Oradea,  Romania}
\affiliation{MTA-BME Quantum Dynamics and Correlations Research Group, Budapest University of Technology and Economics,  M\H uegyetem rkp. 3., H-1111 Budapest, Hungary}
\author{Gergely Zar\'and}
\affiliation{Department of Theoretical Physics, Institute of Physics, Budapest University of Technology and Economics, M\H uegyetem rkp. 3., H-1111 Budapest, Hungary}
\affiliation{MTA-BME Quantum Dynamics and Correlations Research Group, Budapest University of Technology and Economics,  M\H uegyetem rkp. 3., H-1111 Budapest, Hungary}

\begin{abstract}
We study the Coulomb glass emerging from the interplay of strong interactions 
and disorder in a model of spinless fermions  on the Bethe lattice.  In the 
infinite coordination number limit, strong interactions induce a metallic 
Coulomb glass phase with a pseudogap structure at the Fermi energy. 
Quantum and thermal fluctuations  both melt this glass and induce a disordered 
quantum liquid phase.    We combine self-consistent   diagrammatic perturbation theory with continuous time quantum 
Monte-Carlo simulations to obtain the complete  phase diagram 
of the electron glass,  and to characterize its 
dynamical properties in the quantum liquid,  as well as in the  
replica symmetry  broken glassy phase.   Tunneling spectra  display 
 an Efros-Shklovskii  pseudogap  upon decreasing 
temperatures, but the density of states remains finite at the Fermi energy  
due to residual quantum fluctuations.  Our results  bear relevance to the 
metallic glass phase  observed in Si inversion layers.
 \end{abstract}

\maketitle


\section{Introduction}

Describing the  localization of disordered electrons in the presence of long-ranged Coulomb interactions 
and the melting of the Coulomb glass due to quantum-fluctuations represent some of the most 
challenging unsolved and elusive problems  in modern condensed matter physics~\cite{PollakBook}.  
In the absence of interactions, disorder tends to suppress  quantum fluctuations, 
and leads to Anderson localization~\cite{MirlinEvers,Kramer}. 

The presence of interactions, however, changes the structure of localization  transition entirely: unscreened 
Coulomb interactions lead to stronger and stronger anomalies on the metallic side 
as one approaches the phase transition~\cite{Altshuler1980}, 
amount in the formation of curious spin fluctuations~\cite{Sachdev_Bhatt}, and lead to 
the emergence of the Coulomb gap~\cite{Efros_1975, AminiKravtsovMuller_2014} on the insulating side, accompanied by glassy 
dynamics and memory effects~\cite{DraganaPopovic_memory,OregImry_review, Pollak2002, Ovadyahu2007}.  
A major step towards understanding this quantum 
phase transition has been made by A. M. Finkel'stein,
who developed a scaling theory in the presence of Coulomb interactions and weak disorder~\cite{Finkelstein1984}. 
Certain implications of this scaling theory regarding the critical behavior have been verified experimentally~\cite{Zeches977}, 
but a perturbative scaling theory leaves the structure of the localized phase unrevealed,  and has little to say about properties 
of the localized phase such as the formation of the pseudogap or the glassy structure of the localized phase, not to mention
 the connection with  many-body localization~\cite{MBL_review, Mirlin.2018}. 
 
The influence of  quantum tunneling   on the Coulomb gap has been 
addressed initially by means of numerical approaches. 
A configuration interaction approximation  based computation 
predicted a  considerable reduction of the width of the  Coulomb gap~\cite{Vignale1987}, while 
 Hartree-Fock calculations predicted   a modification of  the structure of the classical Efros-Shklovskii 
  pseudogap
  close to the Fermi surface~\cite{Vojta1997, Vojta1998}.

\begin{figure}[b!]
	\includegraphics[width=0.50\columnwidth]{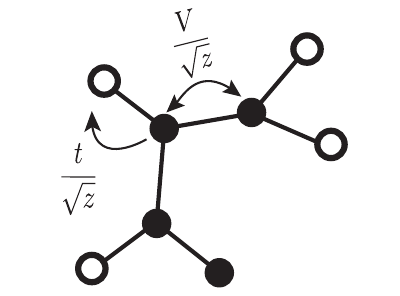}
	\caption{ Sketch of the mean field electron glass model, Eq.~\eqref{eq:Ham}. 
	Electrons move on a disordered Bethe lattice of  coordination number $z$, can hop between neighboring sites, 
	and interact through a Coulomb interaction with nearest neighbors.}	
\label{fig:Bethe_lattice}
\end{figure}

A major step towards  understanding the quantum melting of the Coulomb glass is the construction 
of a solvable mean field theory, similar to the  Sherrington-Kirkpatrick (SK) model, the  
 standard mean field model of classical and quantum spin glass transitions~\cite{SK_book, Binder1986,Parisi_book,Fisher_book}.
Such a mean field model, accounting for most essential properties of the Coulomb glass phase except Anderson localization,   has been proposed by Pastor and Dobrosavljevi\' c in their seminal work~\cite{Pastor1999}, 
possibly inspired by  the extended dynamical mean field  approach applied  to clean correlated systems~\cite{SachdevYe1993,Si1996,Chitra2000}. 
 In the spinless version of their model,  electrons move on a Bethe lattice of coordination
$z\to\infty$, experience some onsite disorder, $\varepsilon_i$, 
and interact with each other through a repulsive nearest neighbor interaction, $V_{ij}= V/\sqrt{z}$, 
mimicking the long-ranged Coulomb interaction (see Fig.~\ref{fig:Bethe_lattice}),
\begin{equation}
\label{eq:Ham}
\hat{H}=- \dfrac{t}{\sqrt{z}} \sum_{\langle i,j\rangle}\left(\hat{c}^\dagger_i\hat{c}_j+{\rm h.c}\right)
+\dfrac{V}{\sqrt{z}}\delta\hat{n}_i\delta\hat{n}_j
+\sum_i\varepsilon_i\delta\hat{n}_i\;.
\end{equation}
Here $\delta\hat{n}_i=\hat{c}_i^\dagger\hat{c}_i-1/2$ denotes deviations from half-filling, and 
the levels $\varepsilon_i $ are drawn from a Gaussian distribution, $P(\varepsilon)\sim e^{- \varepsilon^2 / (2 W^2)}$.
In the rest of the paper, we shall refer to this model as the \emph{disordered  $t-V$ model}.

As shown in Ref.~\cite{Pastor1999}, even though the interaction is uniform, 
spontaneous density fluctuations lead to the emergence of a glass phase,  and 
 the model \eqref{eq:Ham}  maps onto the  Sherrington-Kirkpatrick model
in the absence of quantum tunneling, $t=0$.  We emphasize that this transition is \emph{structural} in the sense
 that it  takes place even in the absence of disorder, $W\to 0$, and is driven by interactions rather than disorder. 
Here, in contrast to the SK model,  frustrations  do not originate  from a frustrated interaction:
rather,  a fluctuation in the 
occupation of some levels  creates a "frustration by choice", leading to the glass transition. 

Later works  revealed a number of key properties of the disordered  $t-V$ model, Eq.~\eqref{eq:Ham}. 
The quantum critical behavior has been analyzed for small disorder in terms of a Landau theory~\cite{Landau1,Landau2},
following a line similar  to the work of Read, Sachdev, and Ye~\cite{ReadSachdevYe95}, and 
it has been argued  that for finite coordination numbers, $z$, a glassy metallic 
 phase should separate  the glassy insulating phase from the  disordered Fermi liquid~\cite{DobrosavljevicTanaskovicPastor2003}, 
as observed on low mobility Si inversion layers~\cite{Popovic2002}.

Nevertheless, in spite of all these achievements and efforts, a complete  solution of  \eqref{eq:Ham}
is still missing, even in the mean field limit, $z\to\infty$.
Here we attempt to give such an accurate and extensive numerical solution of the disordered $t-V$ model in the $z\to \infty $, mean field 
limit. In addition to determining the complete phase diagram, the distribution of local levels, the order parameter, 
{the free energy}, and the {entropy}, we also determine 
the spectral properties and the tunneling spectra of the electrons, as well as their 
scaling properties away from the critical point.  

The solution of  \eqref{eq:Ham} represents a quite challenging task: to enter the glassy phase and capture the formation of the pseudogap, 
we must allow for \emph{complete} replica symmetry breaking, –– accounting for the distribution of local 
(renormalized) energy levels, –-  and, at the same time, we must solve an \emph{ensemble} 
of quantum impurity problems coupled self-consistently back to the spin glass order parameter~\cite{Ioffe2004,Pankov2005,Muller}. 
This route has been followed in Ref.~\cite{Andreanov2012} to study the glassy 
phase of the  – somewhat simpler – transverse field Sherrington-Kirkpatrick model.
Here we derive the appropriate equations for the mean field Coulomb glass model 
by using a path integral formalism, and solve the mean field theory numerically.

\begin{figure}[t!]
	\includegraphics[width=0.95\columnwidth]{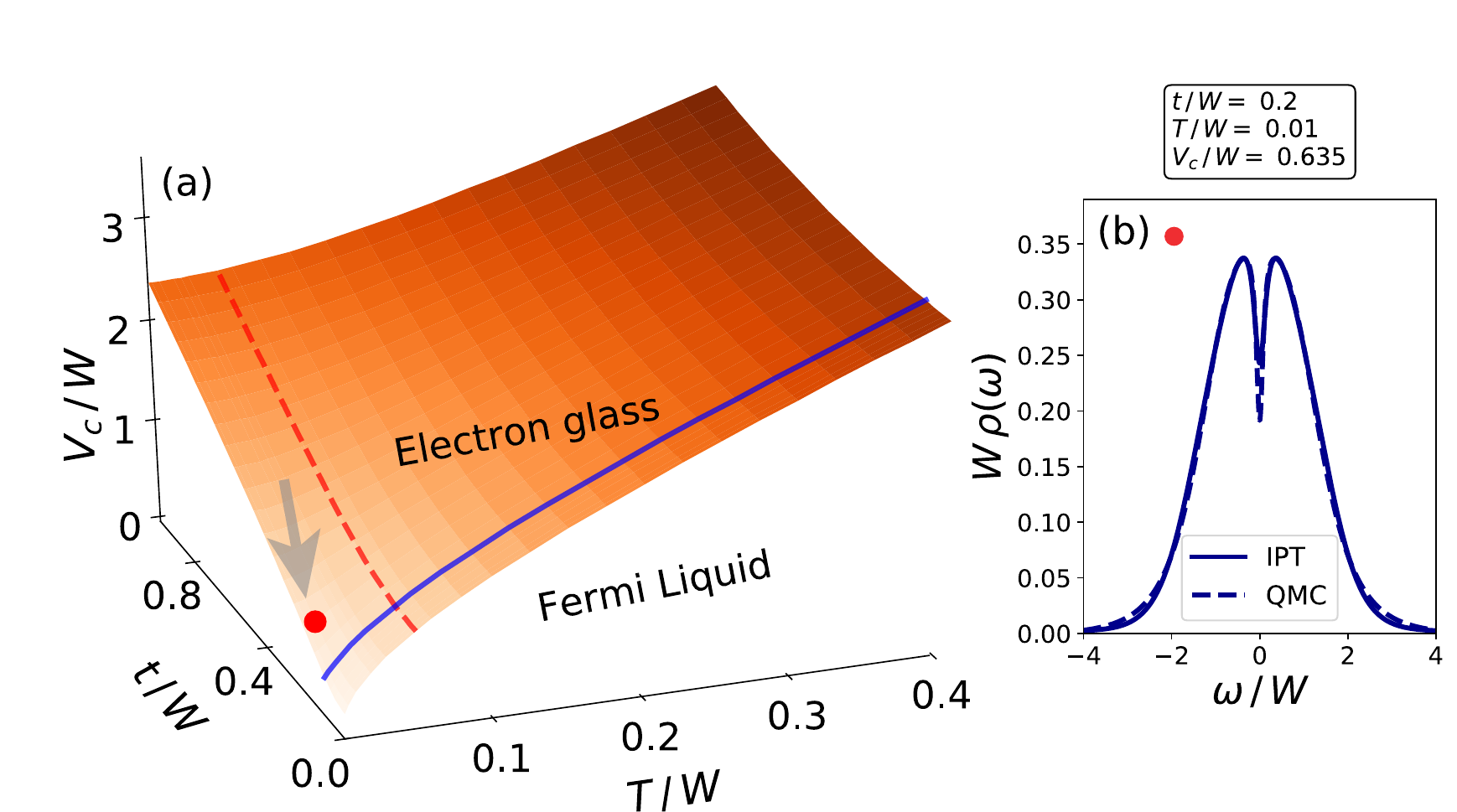}
	\caption{ (a)  Boundary separating the Fermi liquid  and the electron glass phases. 
	Cuts along the solid and dashed lines are presented in Fig.~\ref{fig:phase_diagram_cuts}.	
(b) Spectral function computed at the quantum phase transition point, indicated in panel (a) by the arrow. 
A  correlation hole preceding the pseudogap structure starts to form  already  at the phase boundary.  
}
\label{fig:phase_diagram}
\end{figure}

We  apply two different numerical methods: In the Fermi liquid phase, 
we use an extended  continuous  time Quantum Monte-Carlo (CTQMC) dynamical mean field approach.
This method provides us the numerically exact, self-consistent solution, however, 
is numerically demanding, and  is only appropriate to give us a solution at 
a relatively small number of points in parameter space. 
We therefore combine this approach with an Iterative  Perturbation Theory (IPT), similar 
in spirit to  the one used to describe the Mott transition in a  pioneering work by  A. Georges~\emph{et al.}~\cite{IPT}.
A combination of these two approaches allows us to obtain a coherent  picture, 
summarized in Fig.~\ref{fig:phase_diagram}.

For convenience, we measure all energy scales  in Fig.~\ref{fig:phase_diagram} 
in the disorder strength, $W$. Phase boundaries in Fig.~\ref{fig:phase_diagram}.a correspond to
replica symmetry breaking. The electron glass forms at 
large interactions, and  is destroyed both by thermal ($\sim T$) and by quantum ($\sim t$) fluctuations. 
Typical spectral densities are presented in Fig.~\ref{fig:phase_diagram}(b) at 
a transition point, where $T\ll t$, and therefore quantum fluctuations drive the quantum glass to quantum liquid 
phase transition. A marked correlation hole structure starts to form already at the critical point.
This anomaly gradually develops into a pseudogap that gets deeper and deeper as we enter the glass phase, 
but the density of states remains finite at the Fermi energy in the glass phase for any finite  quantum tunneling, 
even in the $T=0$ temperature limit. This is a peculiarity  of the exactly solvable mean field limit $z=\infty$, where no Anderson localization 
takes place.  The glass state we find is therefore identified as a \emph{metallic} (spinless) \emph{electron glass}, 
observed in several experiments~\cite{Popovic2002, Popovic2012, Moon2018}.  
A similar glassy metallic phase has been predicted to emerge in itinerant fermionic systems 
with cavity mediated long-range interactions, based on a replica symmetric effective field theory approach~\cite{Muller2012LongRange}.


The rest of the paper is organized as follows. 
In Sec.~\ref{sec:mf},
 we introduce the mean field equations through an intuitive cavity approach as well as the more technical replica method,  and
  then formulate the resulting self-consistency equations for the replica symmetric solution and for full replica symmetry breaking. Sec.~\ref{sec:numerics} is devoted to the more technical 
  aspects of the solution. There we discuss the faster but approximate iterative perturbation theory (Sec. ~\ref{subsec:ITP})
  and the exact but numerically more demanding continuous time quantum Monte Carlo (CTQMC) method (Sec. ~\ref{subsec:QMC}). 
  In the rest of the paper we present  our results.  The spectral function in the replica symmetric Fermi liquid phase as well as the phase boundary of  the electron glass phase are presented in  Sec.~\ref{sec:boundary}. 
 Sec.~\ref{sec:rsb} is devoted to the glassy phase: we discuss there the evolution of the order parameter, 
 the distribution of Hartree energies, and  the properties of the spectral function deep in the glassy phase.
  We discuss the thermodynamics of the mean field Coulomb glass model in Sec.~\ref{sec:thermo}.    
  Our main findings are summarized in Sec.~\ref{sec:discuss}. 
 Technical details on the CTQMC simulations and on the numerical solution  in the replica symmetry broken phase
 are relegated to appendices.
 
\section{Mean field equations}\label{sec:mf}

The mean field equations of  disordered $t-V$  model \eqref{eq:Ham} have been derived for $z\to\infty$ using the replica method 
in Ref.~\cite{Pastor1999}.  To our knowledge, these equations have, however, never been solved before 
in their full power. In fact, to obtain a full solution and to capture the formation of the 
Coulomb gap, one must consider the  structure of \emph{full} replica symmetry breaking, just as for  
the Sherrington-Kirckpatrick model~\cite{Rizzo2002_SK,Oppermann2005_SK,Pankov2006_SKmodel,Muller}. 

Before we discuss the quite technical replica method, let us start with a cavity consideration. 
This allows one to  understand the ultimate structure of the equations to be solved. 

\subsection{Cavity approach and effective local action}\label{subsec:cavity}

Let us focus on site $i=0$, and write the action corresponding to the Hamiltonian \eqref{eq:Ham} 
as follows: 
\begin{widetext}
\begin{align}\label{eq:S}
S & = \int_\tau  {\overline c}_{0\,\tau}(\partial_\tau + \epsilon_0) c_{0\,\tau}
+ \frac{V}{\sqrt{z}} {\sum_{i}}^\prime \int_\tau \delta n_{0,\tau} \delta n_{i,\tau}
- \frac{t}{\sqrt{z}} {\sum_{i}}^\prime \int_\tau ( {\overline c}_{0\,\tau}{c}_{i\,\tau} + h.c.) + S_{i\ne0}
\;.
\end{align}
Here we used the shorthand notation, $\int_\tau = \int_0^\beta \dtau$, with $\beta$ the inverse temperature, and have
 separated those pieces, which involve site $0$.  Primes indicate summations over nearest neighbors only. We can then formally expand 
 $e^{-S}$, and integrate out all $i\ne 0$ sites to obtain an effective action for site 0, 
 \begin{align}\label{eq:S0eff}
S_0^{\rm eff}  &=
\int_\tau  \overline{c}_{0\,\tau}(\partial_{\tau} + \epsilon_0) {c}_{0\,\tau}
- t^2 \int_\tau \int_{\tau'}    {\overline c}_{0\,\tau} {c}_{0\,\tau'} \;  \frac 1 z {\sum_{i}}^\prime   \langle c_{i\,\tau} {\overline c}_{i\,\tau'} \rangle_{\rm cav} 
+ \frac{V}{\sqrt{z}}  {\sum_{i}}^\prime \int_\tau \delta n_{0,\tau} \langle \delta n_{i,\tau}\rangle_{\rm cav}\nonumber
\\
&- \frac{V^2}{2}    \int_\tau \int_{\tau'}   \delta n_{0,\tau} \delta n_{0,\tau'}  
\frac 1 z {\sum_{i}}^\prime 
 \langle  \delta n_{i,\tau} \delta n_{i,\tau'} \rangle_{\rm cav} 
\; + \dots.
\end{align}
Here  the $\langle\ldots\rangle_{\rm cav}$ denote cavity averages, i.e. averages computed in the 
absence of site $i=0$.
Higher order contributions, that are not displayed,  vanish in the $z\to\infty$ limit on the Bethe lattice. 
The third term in this expansion represents a random chemical potential, arising from charge fluctuations at neighboring sites. 
We can rewrite the above action as 
 \begin{align}
 \label{eq:S0eff2}
S_0^{\rm eff}  = &
 \int_\tau \int_{\tau'}  \Big\{   \overline{c}_{0\,\tau} \big(  
\delta(\tau-\tau')(\partial_{\tau} + {\tilde \epsilon}_0) - t^2 G(\tau-\tau')\big) {c}_{0\,\tau'} \; 
- \frac{V^2}{2} \chi(\tau-\tau')   \delta n_{0,\tau} \delta n_{0,\tau'}   \Big\}
\;,
\end{align}
\end{widetext}
with $G$ and $\chi$ denoting a nearest neighbour  average over cavity Green's functions 
and dynamical susceptibilities, and 
the random field ${\tilde \epsilon}_0$ incorporating charge fluctuations on neighbouring sites into the bare local 
field, $\epsilon_0$. 
Since the presence of site 0 induces a perturbation of order $\sim 1/\sqrt{z}$ on its nearest neighbors, 
$G$ and $\chi$ can be replaced by the \emph{lattice average} of the  local Green's function and dynamical charge susceptibility, 
respectively.  Would we know the distribution of ${\tilde \epsilon}_0$, $\widetilde P(\tilde \epsilon)$, we could 
replace these spacial averages by an average over $\tilde \epsilon$. We could
thus solve 
the action \eqref{eq:S0eff2} for $G_{\tilde \epsilon}(\tau)$ and $\chi_{\tilde \epsilon}(\tau)$, 
and obtain    $G(\tau)$ and $\chi(\tau)$ by averaging over $P(\tilde \epsilon)$, thereby closing  a dynamical mean field theory (DMFT) cycle.

Unfortunately, it is not so simple to obtain $\widetilde P(\tilde \epsilon)$. The difficulty is related to spontaneous 
symmetry breaking. Even for a given set of the on site energies, $\{\epsilon_i\}$, each 'leg' attached to the cavity 
has namely \emph{ many} symmetry broken states. We should pick a symmetry broken charge 
distribution on each of these legs. However, we \emph{cannot} choose these independently of each other, since 
the central site $i=0$ creates correlations between the legs. 
Adding/removing the site 0 induces  namely a correlated charge shift  of order $\sim1/\sqrt{z}$ on neighboring 
charge distributions. This,  in turn, amounts in a change of ${\cal O}(1)$ in the value of $\tilde \epsilon_0$ and,  
more importantly, correlates the occupation of neighboring sites  through charge fluctuations at site $i=0$. 
The situation is quite similar to that of the ferromagnetic phase of an Ising magnet on the Bethe lattice, where the direction 
of magnetization on each leg  gets correlated through the central site.  

The appropriate distribution $\widetilde P(\tilde \epsilon)$ follows from \emph{stability criteria}, 
usually formulated in terms  of the replica method, discussed in the next subsection. 
We shall also follow this – somewhat formal –  route to determine  $\widetilde P(\tilde \epsilon)$.

\subsection{Replica approach}\label{subsec:replica}

The action  $S_0^{\rm eff}$  in Eq.~\eqref {eq:S0eff2} can also be obtained via the replica trick, 
whereby we  express the logarithm of the partition function as 
\begin{equation}\label{eq:rep}
\log Z =\lim_{n\rightarrow 0}\dfrac{Z^n-1}{n}\;.
\end{equation}
We therefore take $n\to 0$ copies of the Hamiltonian, integrate out the Gaussian disorder 
and the fermions  at all sites excepting the central one. The latter step becomes 
simple in the $z\to \infty$ limit, where a systematic $1/z$ cumulant expansion  leads  to a simple 
(extended) dynamical mean field theory structure~\cite{DMFT_review} with the effective action~\footnote{Note that Eq. \eqref{eq:Srep} neglects certain (temperature dependent) constant terms resulting from integrating out fermions at all lattice sites except one. These terms, not involving the Grassmann variables $c^{\,a}_\tau$ and $\overline c^{\,a}_\tau$, do not enter the correlators of the local effective model, however, they will become important when we study  thermodynamic properties of the full lattice Hamiltonian.}
\begin{widetext}
\begin{align}\label{eq:Srep}
S_{\rm rep}=\int_0^\beta d\tau\int_0^\beta d\tau^\prime&\Bigg\lbrace\sum_{a=0}^n\left(\overline{c}_{\,\tau}^{\,a}\left[\delta(\tau-\tau^\prime)\partial_{\tau^\prime}-t^2 G(\tau-\tau^\prime)\right] {c}_{\,\tau'}^{\,a}-\dfrac{V^2}{2}\chi(\tau-\tau^\prime)
\delta n_{\,\tau}^{\,a} \; \delta n_{\,\tau'}^{\,a} \right)\nonumber\\
&\qquad-\dfrac{1}{2}\sum_{a\ne b }^n\ V^2 Q^{\,ab} \delta n_{\,\tau}^{\,a} \delta n_{\,\tau'}^{\,b} 
 -  \dfrac{1}{2} \sum_{a,b=0}^n  W^2    \delta n_{\,\tau}^{\,a} \delta n_{\,\tau'}^{\,b} 
\Bigg\rbrace.
\end{align}
The action \eqref{eq:Srep} is supplemented  by the self-consistency conditions, 

\begin{align}\label{eq:consistency2}
G(\tau-\tau^\prime)  =\left\langle c^{\,a}_\tau\, \overline c ^{\,a}_{\tau^\prime}\right\rangle_{_{S_\text{rep}}},\quad\quad
\chi(\tau-\tau^\prime)&=\left\langle\delta n^a_\tau\,  \delta n^a_{\tau^\prime}  
\right\rangle_{S_{\rm rep}},\quad\quad
Q^{a\ne b}=
\large\langle\delta n^a_\tau  \,  \delta n^b_{\tau^\prime}  \large\rangle_{S_\text{rep}}\,.
\end{align}
\end{widetext}
Disorder appears at this point only  through the term $\sim W^2 $, coupling (correlating) different 
replicas, and the off-diagonal structure 
of the glass order parameter, $Q^{a\ne b}$, capturing density fluctuation correlations between different replicas subject 
to the same disorder.  
 This replica-replica coupling in $S_{\rm rep}$ may lead to spontaneous  
 replica symmetry breaking, characteristic to the glassy phase, and 
signaling that  replicas break ergodicity individually and 
differently. In the next subsection we first address the simpler replica 
symmetric symmetric solution, before sketching the procedure in   the regime of full replica symmetric breaking
in subsection \ref{subsec:rsb}. For more technical details, we refer the reader to 
 Appendix ~\ref{app:rsb}.

\subsection{The replica symmetrical self-consistency equations}\label{subsec:rs}

In general, the non-trivial replica structure of $Q^{ab}$ leads to difficulties when taking the limit, $n\rightarrow 0$. 
The equations  simplify, however,  considerably  in the (non-glassy)  replica symmetrical phase, 
where all replicas behave in the same way, and  $Q^{a\ne b} = Q_\text{RS}$ for all $a\ne b$. 
This phase is identified as a disordered  Fermi liquid phase~\cite{DobrosavljevicTanaskovicPastor2003}.

 In this case,  we can decouple 
the off-diagonal part of the  last term of the effective action \eqref{eq:Srep} with a Hubbard-Stratonovitch 
field, $\etilde$, leading to the local  effective action, 
\begin{eqnarray}
\label{eq:Seff}
S_\etilde&=\int_\tau \int_{\tau^\prime}
\Big\lbrace 
\overline{c}_\tau\Big[ 
\delta_{\tau,\tau^\prime} \left[\partial_{\tau^\prime} +\etilde \right]-t^2 \; G(\tau-\tau^\prime)
\Big]\; c_{\tau^\prime}  \nonumber\\
- &\dfrac{V^2}{2 }\, \large(\chi(\tau-\tau^\prime)- Q_\text{RS}\large)\; \delta n_\tau \;\delta n_{\tau^\prime} 
\Big\rbrace 
-\dfrac{\beta\;\widetilde \varepsilon}{2}, 
\end{eqnarray}
with the Hubbard-Stratonovic field $\tilde \epsilon$ a  Gaussian variable of distribution
$\Ptilde_\text{RS}(\etilde)\sim  \exp\big(-\etilde^2/(W^2 + V^2Q_\text{RS})/2\big)$. The 
 self-consistency equations \eqref{eq:consistency2}  are now replaced by the conditions, 
\begin{equation}
\left\{\begin{array}{c}
       G(\tau) \\
{\chi}(\tau)
        \end{array}\right\}
        = \int d\etilde \,  \Ptilde_\text{RS}(\etilde,\; Q_\text{RS} ) \,
      \left\{\begin{array}{c}  
        {G}_\etilde(\tau\,)\\
    { \chi}_\etilde(\tau\,)
        \end{array}\right\}\,,
        \label{eq:average}
\end{equation}
and $Q_\text{RS}$ is also determined selfconsistently by  
\begin{equation}
Q_\text{RS} = \overline{ \langle\delta n\rangle^2} =  \int \text{d}\etilde\, \Ptilde_\text{RS}(\etilde,\; Q_\text{RS} ) \; {\langle \delta n\rangle_{\etilde}}^{2}\;. 
\label{eq:QRS}
 \end{equation}
 with ${G}_\etilde(\tau\,)$,  ${ \chi}_\etilde(\tau\,)$, and  $ \langle\delta n\rangle_{\etilde}$ computed by the effective (local) action, 
Eq.~ \eqref{eq:Seff}. 
 
In the replica symmetrical case, we  thus converted the problem into an  \emph{ensemble} of local, self-interacting 
fermion levels. The width of the distribution of the level $\tilde\epsilon$ as well as the fermion's self-energy
($\sim t^2 G(\tau)$) and its self-interaction ($\sim V^2 [\chi(\tau)-Q_\text{RS}]$) must all be determined self-consistently. We defer discussing the numerical solution of this ensemble of local actions, i.e. the computation of the quantities 
 ${G}_\etilde(\tau\,)$,  ${ \chi}_\etilde(\tau\,)$, and  $ \langle\delta n\rangle_{\etilde}$,  to Sec.~\ref{sec:numerics}.

 Remarkably, the local action has \emph{exactly}  the same structure as  \eqref{eq:S0eff2}. 
However, the replica approach also provides us the self-consistent 
distribution function $\tilde P(\etilde)$:  in the replica symmetrical Fermi liquid phase, 
the "Hartree field" distribution  retains the Gaussian structure of the bare disorder $\varepsilon_i$, 
and interactions only renormalize the variance of the effective field.

Importantly, in the classical limit, $t=0$, 
we can set 
 $ G\to 0$ {when we determine the occupancy, and 
 $\delta n_\tau \;\delta n_{\tau^\prime} = 1/4$. Then we
   simply obtain 
$\langle \delta n\rangle_{\etilde}=-\tanh (\etilde/(2T))/2$}. Eq.~\eqref{eq:QRS} then just becomes essentially the
self-consistency equation of the Sherrington-Kirkpatrick (SK) model in case of replica symmetry~\cite{SK_book, Muller}. 
The mean field Coulomb glass problem is  thus equivalent to the SK model in the 
classical limit, as pointed out in Ref.~\onlinecite{Pastor1999}. 
However,  contrary to the SK model, where the replica symmetrical solution {with $Q_{RS}\neq 0$}
is intrinsically unstable, here replica symmetry is stabilized by finite disorder as well as finite 
quantum fluctuations, and a valid replica symmetric phase exists.

\subsection{Full replica symmetry breaking}\label{subsec:rsb}

In the electron glass phase, replica symmetry is fully broken.  Fortunately, 
the  construction of the previous section can be generalized to incorporate full replica symmetry breaking, 
thereby yielding a complete description of the glassy phase as well. Although  derivations may seem 
cumbersome,  the interpretation of the final results is  relatively straightforward. 

The local effective action Eq.~\eqref{eq:Seff}, supplemented by the self-consistency conditions Eq.~\eqref{eq:average},
remains unaltered, except for changing $Q_{\rm RS}\rightarrow Q_{aa}$, expressing that electrons at each site experience a different "Hartree field", 
$\varepsilon_i \to \widetilde \varepsilon_i$, due to the conspiracy  of random onsite energies and 
nearest neighbor Coulomb interactions. Only the "Hartree field's" distribution $\Ptilde(\etilde)$ acquires a more complicated, 
non-Gaussian structure, that must  be determined self-consistently together with the average propagators and susceptibilities  
(see Appendix~\ref{app:rsb}).  

The solution of the ensemble of effective actions,  Eq.~\eqref{eq:Seff}, has to be carried out exactly the same way as in the replica symmetrical 
phase.  Only the last,  least intuitive step of this derivation,  the determination of the distribution of the renormalized "Hartree" 
energies  $\widetilde P(\etilde)$ is much more difficult. The derivation of this distribution 
 follows similar lines as the solution of the classical spin-glass problem~\cite{SK_book}, apart from the 
 fact that here we need to work with the quantum action.  
 
We parametrize $Q^{ab}$ using Parisi's variables~\cite{Parisi_book} as a function 
$Q(x)$, with the  
replica variable $x\in [0,1]$ parametrizing deeper and deeper layers of replica symmetry 
breaking. In the replica symmetrical phase $Q(x)$ remains constant, while in the glassy phase $Q(x)$ is no longer
constant,  $Q(x)\neq \mathrm{constant}$. This property allows us  to detect the boundary of the glassy phase.
Alternatively, we can determine the boundary by using
 the stability condition \eqref{eq:stabcond}, discussed later in Section~\ref{sec:boundary}.
 For a complete solution, we  need to generate a \emph{family} of  effective 
actions, parametrized by $x$. 
Physical quantities at different layers are related by so-called "flow equations". 
 The final structure of these latter is  outlined in Appendix~\ref{app:rsb}.  
 
 Fortunately, the flow equations are decoupled from the quantum solution 
 in the sense that the quantum problem 
 only provides  boundary conditions for them.
 In fact, as input 
 one only  needs the  (negative) free energy of the  embedded level,
$\Phi_\etilde(T)\equiv k_B T\,\ln\,Z_\etilde$, defined by 
\begin{equation}
e^{\,\beta \Phi_{\etilde}( T)} \equiv  \int \mathcal{D}\overline c \, \mathcal{D}c\;e^{-S_\etilde[\,\overline c,c\,]}\;. 
\label{eq:philoc_gen}
\end{equation}
The solution of the flow equations then provides the renormalized 
distribution, $\tilde P(\etilde)$,  and the order parameter $Q(x)$.

\vskip0.5cm
\section{Solving the mean field equations}
\label{sec:numerics}

To obtain the full solution of the disordered $t-V$ model, we have to solve an ensemble of local actions, Eq.~\eqref{eq:Seff}, i.e., compute the quantities 
 ${G}_\etilde(\tau\,)$,  ${ \chi}_\etilde(\tau\,)$, and  $ \langle\delta n\rangle_{\etilde}$. In the replica symmetric phase, we simply 
 iterate the self-consistency equations Eqs.~\eqref{eq:average} and \eqref{eq:QRS}, whereas in the presence of full replica symmetry breaking Eq.~\eqref{eq:QRS} is replaced by more complicated flow equations (see Appendix ~\ref{app:rsb}). To treat the action in 
 Eq.~\eqref{eq:Seff}, we employed two different methods: for a fast but approximate solution, 
we used Iterative Perturbation Theory (IPT), which allowed us to get a complete solution  in the 
replica symmetric Fermi liquid phase as well as to access  the glass phase. 
To complement this approach, we have also obtained a "numerically exact" solution
by the Continuous Time Quantum Monte Carlo method (CTQMC). Since the CTQMC method is computationally very demanding, we only applied it in the replica symmetric phase, where we used it to obtain   reference 
solutions at many points in the parameter space, and also to verify the  phase boundaries. 

\subsection{Iterative Perturbation Theory}\label{subsec:ITP}

\begin{figure}[b!]
	\includegraphics[width=0.85\columnwidth]{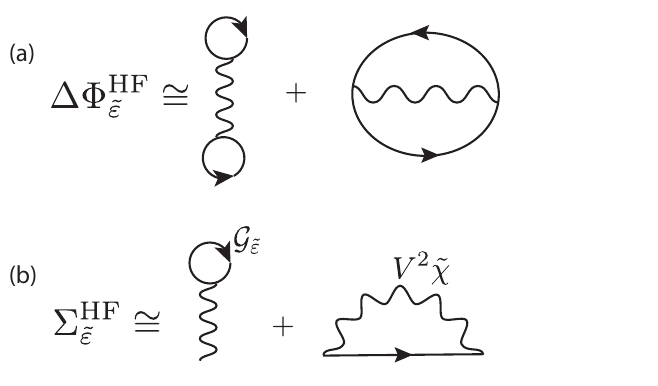}
	\caption{ (a)  "Hartree-Fock" local free energy contributions and  (b) 
	 corresponding self-energy diagrams.  (Counter-term diagrams are not shown.) Wavy lines represent the effective interaction, 
	{$V^2\,\widetilde\chi(\tau-\tau^\prime)$}, while continuous lines stand for the 
	unperturbed  local propagator, $\mathcal{G}_\etilde(\tau) = \langle c(\tau) \bar c (0)\rangle_{\rm loc}^{(0)}$, 
	 computed from the non-interacting part 
	of Eq.~\eqref{eq:Seff} . 
	}
	\label{fig:HF_free_energy}
\end{figure}

The effective action Eq.~\eqref{eq:Seff} describes a fermion propagating 
{with the unperturbed propagator $\mathcal{G}_\etilde$ associated with the first term of Eq.~\eqref{eq:Seff},
\begin{equation}\label{eq:Gbare}
\mathcal{G}^{-1}_\etilde(\tau) = \delta_{\tau,\tau^\prime} \left[\partial_{\tau^\prime}+ \etilde \right]-t^2 \; G(\tau-\tau^\prime),
\end{equation}}
and interacting through the retarded interaction
$$
V^2 \;\tilde \chi(\tau-\tau') \equiv V^2\;\large(\,\chi(\tau-\tau') -  \langle \delta n\rangle ^2\large)\;.
$$

To compute all needed Green's functions and susceptibilities in a
systematic way, it is convenient to formulate the approximation in terms of the  the local (negative) free energy $\Phi_\etilde$, Eq. ~\eqref{eq:philoc_gen}. We express $\Phi_\etilde$  as 
\begin{equation} 
 \Phi_\etilde =  \Phi^{(0)}_{\etilde} +  \Delta\Phi_\etilde\, 
 \label{eq:phi_loc}
\end{equation} 
with the second term accounting for the interaction-induced part of $ \Phi_{\etilde}$, and  $ \Phi^{(0)}_{\etilde} $ 
being the non-interacting free energy, 
\be
 \Phi^{(0)}_{\etilde}  = \dfrac{\etilde}{2}+\dfrac{1}{\beta}{\rm Tr}\log \mathcal{G}_\etilde^{-1}\;.
 \label{eq:phi_loc^0}
\ee
The interacting part $\Delta\Phi_{\etilde}$ can  be considered as a functional of the 
dressed propagator.  Then its functional differential with respect to the dressed propagator 
is just the self-energy.

\begin{figure}[t!]
	\includegraphics[width=0.90\columnwidth]{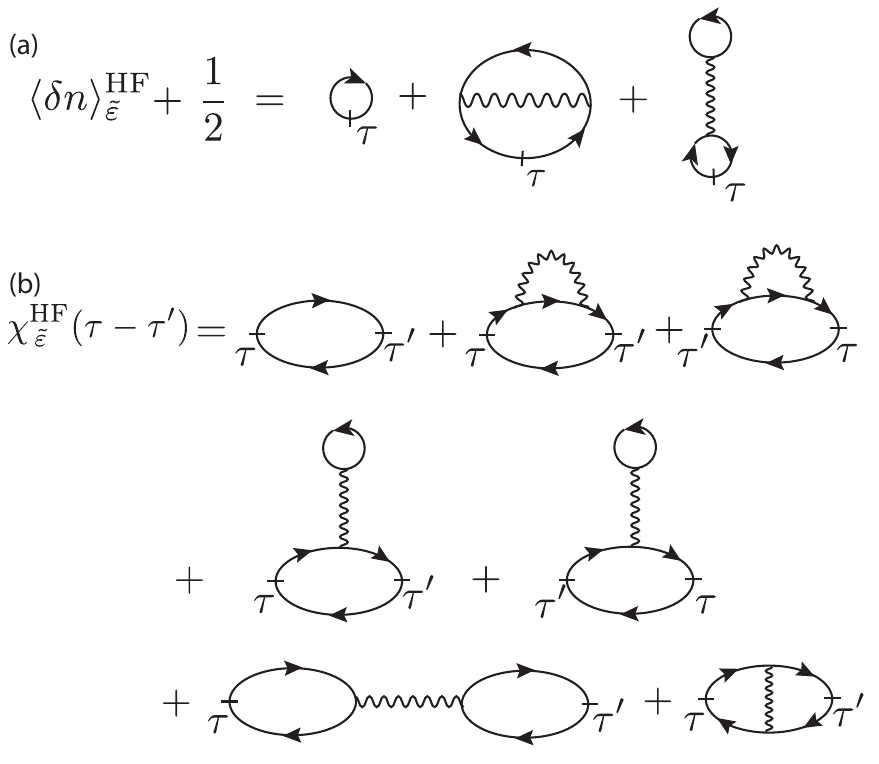}
	\caption{Hartree-Fock diagrams determining (a) the Hartree-Fock occupation, $\langle\delta n\rangle_\text{loc}^\text{HF}$, and (b) the Hartree-Fock response, $\chi_\etilde^\text{HF}(\tau)$.
	Cuts   indicate  functional derivatives,  counter-term diagrams are omitted. 
	}
	\label{fig:chi_HF}
\end{figure}

Within IPT, we simply replace the  local free energy \eqref{eq:phi_loc} by 
the second order perturbative expression,
\begin{align}
  \Delta\Phi^\text{HF}_{\etilde} (\etilde) 
  & =
 \dfrac{V^2}{2} \big(\mathcal{G}_\etilde(0^-)+{1}/{2}\big)^2
\int_0^\beta d\tau\,\tilde{\chi}(\tau)
\label{eq:DeltaPhi_HF}
\\
& - \dfrac{ V^2}{2}\int_0^\beta d\tau\,\tilde{\chi}(\tau)\,\mathcal{G}_\etilde(\tau)\,\mathcal{G}_\etilde(-\tau),
\nonumber
\end{align}
represented by  the free energy diagrams in Fig.~\ref{fig:HF_free_energy}(a) ~\footnote{For clarity, we omitted diagrams incorporating the counter-terms arising from the chemical potential of the half-filled model in Figs. ~\ref{fig:HF_free_energy} and ~\ref{fig:chi_HF}. These diagrams  can be constructed by replacing propoagator loops $\mathcal{G}_\etilde(0-)$ with a constant factor -1/2.}. 
Although not constructed in terms of the full Green's function, we shall also 
refer to this approximation as "Hartree-Fock" approximation, as also inferred by the labels, "HF".
For the self-energy, we use a similar approximation, represented in  Fig.~\ref{fig:HF_free_energy}(b)
\begin{eqnarray}\label{eq:sigma}
\Sigma_{\,\etilde}^\text{HF}(\tau)&= &\delta(\tau)V^2  \big(\mathcal{G}_{\,\etilde}(0^-)+{1}/{2}\big)  
\!\int_0^\beta \!\!d\tau^\prime\widetilde{\chi}_{\,\etilde}(\tau^\prime)
\nonumber \\
&-&V^2\widetilde{\chi}_{\,\etilde}(\tau)\mathcal{G}_{\,\etilde}(\tau)\;.
\end{eqnarray}
These expressions can also be obtained by functional differentiation of \eqref{eq:DeltaPhi_HF} with respect to the 
unperturbed propagators. The term 1/2 originates from normal ordering, and is just the average occupation.

Formally, the occupation  $\langle \delta n\rangle_\etilde$ and for the local compressibility $\chi_\etilde(\tau)$
can be computed  by inserting  a time dependent energy in the action,  $\etilde\to{\widetilde \epsilon}_\tau$, and taking 
 the functional derivatives of $ \Phi_\etilde$ with respect to   $\etilde\to{\widetilde \epsilon}_\tau$.
We use this procedure to obtain the IPT  expressions {for the occupation $\langle \delta n\rangle_\etilde$ and} 
for the local compressibility $\chi_\etilde(\tau)$, consistent 
with the approximations above, by just differentiating $\Phi_\etilde^\text{HF} =  \Phi_\etilde^\text{(0)} + \Delta \Phi_\etilde^\text{HF}$ 
 as 
\begin{equation}
\langle\delta n\rangle_{\etilde}^\text{HF}
=  \beta \left . {\dfrac{\delta \Phi_\etilde^\text{HF}}{\delta\,{\widetilde \epsilon}_\tau}} \right|_{{\widetilde \epsilon}_\tau\to \etilde}
\end{equation}
and
\begin{equation}
\chi^\text{HF}_{\,\etilde}(\tau-\tau^\prime)=\beta\, \left. {\dfrac{\delta^2\Phi_\etilde^\text{HF}}
{\delta\,{\widetilde \epsilon}_\tau\;\delta\,{\widetilde \epsilon}_{\tau^\prime}}} \right|_{{\widetilde \epsilon}_\tau\to \etilde}\;.
\end{equation}
The resulting  expressions 
are quite lengthy, we therefore do not display them here, but the corresponding diagrams, shown  in Fig.~\ref{fig:chi_HF}, 
have a quite transparent structure,  and it is easy to construct the explicit formulas 
from them by  following  standard diagramatic rules. 

In the RS phase, the IPT iteration is then  straightforward. Assuming some ansatz for $Q_\text{RS}$, 
$\chi(\tau)$, and $G(\tau)$ we use Eq.~\eqref{eq:sigma} to compute  $\mathcal{G}_\etilde (\tau)$, 
and the diagrams in Fig.~\ref{fig:chi_HF} to determine $\langle \delta n\rangle_\etilde$
and $\chi_\etilde(\tau)$ for a dense set of energies, $\etilde$. 
We then determine $Q_\text{RS}$,  $\chi(\tau)$, and $G(\tau)$ iteratively by 
means of Eqs.~\eqref{eq:average} and \eqref{eq:QRS}. In the presence of full replica symmetry breaking, the function $Q(x)$ is determined by more complicated flow equations instead of the RS expression ~\eqref{eq:QRS}, but the rest of the iteration loop remains unaltered.

\subsection{Continuous time quantum Monte Carlo }\label{subsec:QMC}



An alternative route to compute $G_{\tilde{\varepsilon}}(\tau)$, $\langle \delta n \rangle_{\tilde{\varepsilon}}$, and $\chi_{\tilde{\varepsilon}}(\tau)$ within the dynamical mean-field theory is to perform a continuous-time quantum Monte Carlo (CTQMC) computation with the effective local action $S_{\etilde}$ given in Eq.~(\ref{eq:Seff}). 
We use an extension of the hybridization-expansion CTQMC algorithm that can treat retarded interactions in action formalism~\cite{werner-EDMFT, werner-dynamical-screening-1}. In this approach we expand the partition function $ Z_\etilde={\rm Tr}\,{\rm e}^{-S_{\etilde}}$ in the hybridization function $F(\tau-\tau^{\prime}) = t^2\,G(\tau-\tau^{\prime})$ while we treat the level energies $\tilde{\varepsilon}$ and interaction $V$ exactly.

In general, the hybridization-expansion CTQMC method\cite{Werner1, Werner2} relies on the expansion of the partition function $Z_\etilde$ in the hybridization into a series of diagrams and sampling these diagrams stochastically,
where $Z_\etilde$ can be written as a sum of configurations $z_{k}$ with weight $w(z_{k})$ as $Z_\etilde = \sum_{z_{k}}w(z_{k})$.
In the segment picture a Monte Carlo configuration $z_{k}$ with expansion order $k$ is represented by $k$ segments with imaginary time intervals $\{\tau_{1}, \tau^{\prime}_{1}\}, ..., \{\tau_{k}, \tau^{\prime}_{k}\}$ where the particle number is 1, and it is 0 where there is no segment. 

In our case the creation operators $\overline{c}_{\tau_{i}}$ at times $\tau_{i}$ are connected to annihilation operators $c_{\tau^{\prime}_{j}}$ at times $\tau^{\prime}_{j}$ by the hybridization function $F(\tau_{i} - \tau^{\prime}_{j}) $, and the collection of these $k!$ diagrams corresponding to the hybridization lines $F$ is summed up into a determinant of a matrix $\hat{F}^{(k)}$ composed of the hybridization functions.
The weight $w(z_{k})$ is  expressed as $w(z_{k}) = {\rm det}\hat{F}^{(k)}\,w_{\tilde{\varepsilon}}\,w_{\tilde{\chi}}$ where the contributions $w_{\tilde{\varepsilon}}$ and $w_{\tilde{\chi}}$ corresponding to the level energy $\tilde{\varepsilon}$ and the interaction term $V^2\tilde{\chi}$ are given in Eqs.~(\ref{wtildeepsilon}) and (\ref{eq-wchin}) in Appendix~\ref{app:CTQMC}, respectively. For further details, please visit this Appendix ~\ref{app:CTQMC}.

Since the CTQMC method is computationally very demanding, we applied it as a reference point in the RS phase. We used the extended CTQMC impurity solver in the numerical calculations with the combined weight $w(z_{k})$ by means of the Metropolis algorithm to solve the effective local action given in Eq.~(\ref{eq:Seff}) for $G_{\tilde{\varepsilon}}(\tau)$, $\langle \delta n \rangle_{\tilde{\varepsilon}}$, and $\chi_{\tilde{\varepsilon}}(\tau)$ self-consistently in the Fermi-liquid (replica symmetric) phase. 
We proceeded to obtain the self-consistent replica symmetric solution through the following iteration steps:
We may start by an arbitrary ansatz for $G^{[0]}(\tau)$, $\tilde{\chi}^{[0]}(\tau)$, and $Q^{[0]}_{\rm RS}$, for example with the non-interacting Green's function and susceptibilities (for the explicit expressions please visit Appendix~\ref{app:CTQMC}), at the zeroth iteration step. We then compute the quantities $G^{[1]}_{\tilde{\varepsilon}}(\tau)$, $\tilde{\chi}^{[1]}_{\tilde{\varepsilon}}(\tau)$, and $\langle \delta n \rangle_{\tilde{\varepsilon}}$ at the subsequent iteration step with the effective local action, Eq.~(\ref{eq:Seff}), using the CTQMC impurity solver for a wide range of level energies $\tilde{\varepsilon}$. The averaged quantities $G^{[1]}(\tau)$, $\tilde{\chi}^{[1]}(\tau)$, and $Q^{[1]}_{\rm RS}$ are obtained by (numerical) integration over $\tilde{\varepsilon}$ with the distribution $\tilde{P}(\tilde{\varepsilon})$ as it is given in Eqs.~(\ref{eq:average}) and (\ref{eq:QRS}). They are used for the next iteration step, and we repeat this procedure until we reach convergence.

We calculated several points of the phase boundary by CTQMC using the stability condition given in Eq.~\eqref{eq:stabcond} below for various parameter values for $t$, $V$ and $T$, and found excellent agreement between the IPT and CTQMC calculations.
The spectral functions are also compared and found to show similar energy dependence between IPT and CTQMC. However, around zero energy $\omega \sim 0$ difference arises  in the density of states  between the IPT and the numerically exact solution as we approach the glassy phase by increasing the interaction $V$, or decreasing the hopping $t$.

\section{Replica symmetric spectral functions and phase boundary}\label{sec:boundary}

We used both approaches described in the previous section
 to compute the Green's function $G(\tau)$ and the 
susceptibility $\tilde \chi(\tau)$ in the replica symmetric phase.  The average local tunneling density of 
states (DOS) can then be computed from the  Fourier transform  $G(i\omega)$ as 
\begin{equation}\label{eq:rho}
\rho(\omega)=\dfrac{1}{\pi}\,{\rm Im}G(i\omega\to \omega+i\,0^+)\;.
\end{equation}
In the very last step, we have used a Pad\'e construction to carry out 
the analytical continuation.

\begin{figure}[bt]
\includegraphics[width=0.90\columnwidth]{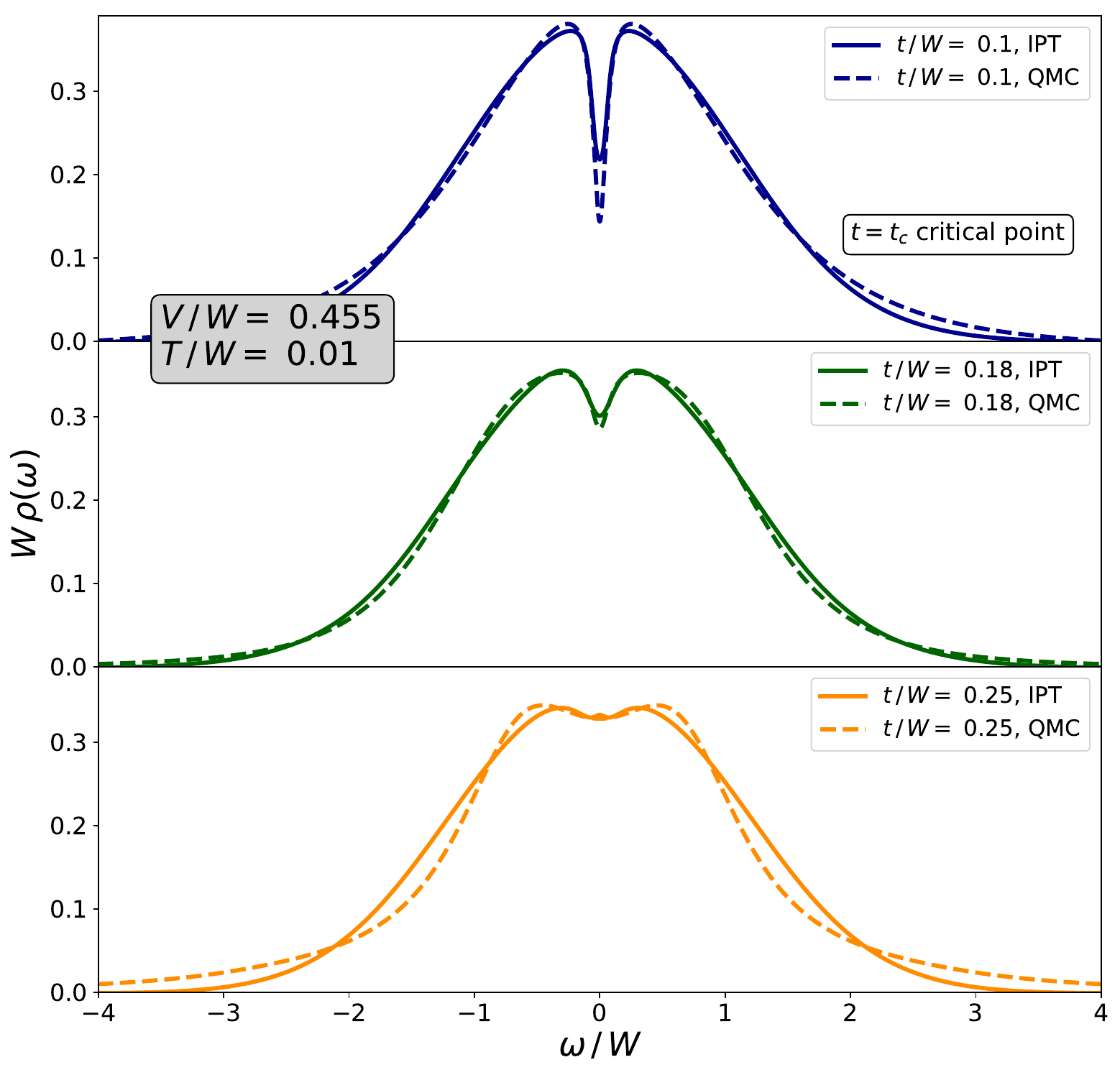}
\caption{Average low temperature density of states in the Fermi liquid phase, as computed 
by CTQMC and by IPT. The density of states develops a remarkable zero bias anomaly 
already in the Fermi liquid, though the distribution of Hartree levels is still featureless. 
Increasing quantum fluctuations wash away this correlation hole.}
\label{fig:Fermi_liquid_spectral_functions}
\end{figure}

\begin{figure*}[bt]
	\includegraphics[width=1.40\columnwidth]{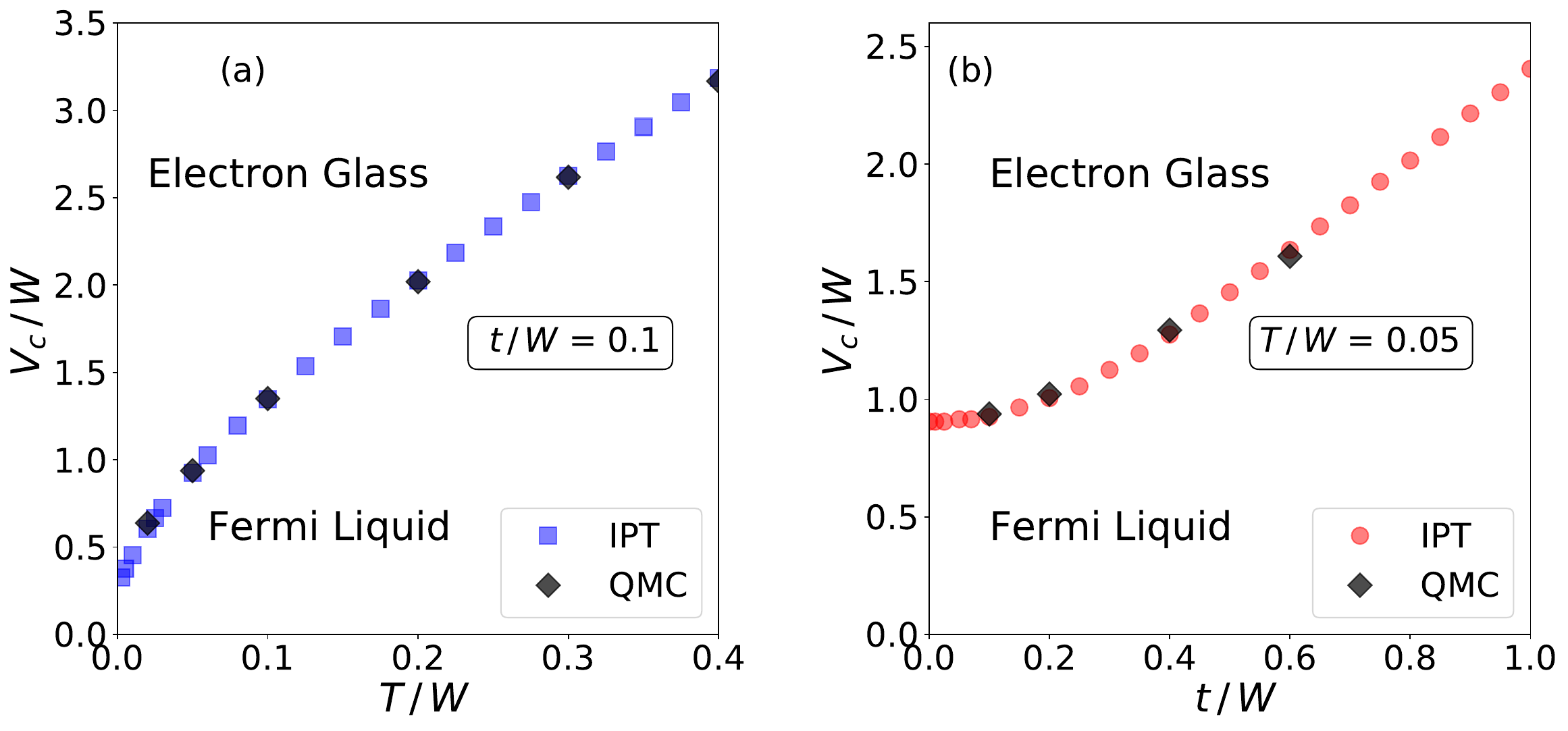}
	\caption{
	\label{fig:phase_diagram_cuts}
	Phase boundary between the electron glass and the Fermi liquid, 
	as computed by continuous time Monte Carlo and IPT along the lines indicated in Fig.~\ref{fig:phase_diagram}.  Interactions lead to the formation of the glass. Quantum fluctuations as well as thermal fluctuations melt  the electron gas.}
\end{figure*}

Fig.~\ref{fig:Fermi_liquid_spectral_functions} shows the spectral functions
in the Fermi liquid phase for a moderate interaction, 
$V/W=0.455$, and quite small temperature, $T/W=0.01$,
as we drive the system closer and closer to the Fermi liquid - electron glass 
quantum phase transition.  For large $t$, quantum fluctuations destroy 
the electron glass, and a dirty Fermi liquid is formed. There the density of states is 
almost featureless.  As we decrease $t/W$, quantum fluctuations get suppressed, and 
a plasma dip structure starts to form in the middle of the band, even though the system still remains in the replica symmetric Fermi liquid phase. 
IPT and CTQMC are in very good agreement, and yield both   very similar structures. 
Differences can  be attributed  to the approximations made within IPT,
 to the limited  CTQMC accuracy,  and to uncertainties related to 
 the analytical continuation. 

 The presence of a plasma dip (or correlation hole) in the Fermi liquid phase reflects short range charge correlations due to the repulsive interactions between neighboring sites. This correlation hole is a manifestation of Onsager's back reaction, and it is not directly 
 related to the Efros-Shklovskii gap of the glassy phase~\cite{Pankov2005,Muller}. 
 Indeed, in the liquid phase, replica symmetry is maintained, implying that the distribution 
 of the renormalized Hartree-Fock levels, $\tilde P(\tilde \epsilon)$ is still 
 a featureless Gaussian, in contrast to the tunneling density of states.

The  boundary of the electron glass phase is determined by a
 stability (marginality)  condition against replica symmetry breaking, ensuring the stability of the solution $Q_{ab}\equiv Q_{\rm RS}$. This  is essentially identical to 
 the stability  condition  appearing in the SK model~\cite{Pastor1999}, 
\begin{equation}
\label{eq:stabcond}
1=V^2 \int\,\text{d}\etilde\,\widetilde P_\text{RS}(\etilde)\,\chi_{\rm stat}^2(\etilde),
\end{equation}
with the static local susceptibility defined as  
$\chi_{\rm stat}(\etilde )\equiv\partial_\etilde \langle\delta n\rangle_\etilde$.

The resulting phase diagram has been presented in Fig.~\ref{fig:phase_diagram}
for a finite disorder, $W$. 
At any temperature and for any hopping $t$, replica 
symmetry is broken at interactions larger than some critical value, $V\ge V_C(T,t,W)$.  
In the classical limit, $t=0$, in particular, an interaction-driven glass phase emerges 
at low temperatures  for small disorder.  Contrary to naive expectations,  strong \emph{disorder  
destroys} the glassy phase, and leads  to a trivial strongly disordered phase without 
replica symmetry breaking (RSB): fluctuations 
of the bare levels $\varepsilon_i$ are so large that each level becomes occupied or unoccupied 
essentially independently, leaving no room to interaction-induced frustration. 
For sufficiently strong interaction, however,  a Coulomb glass phase emerges.

The glass phase can be destroyed not only  by extreme disorder, but also 
by thermal and quantum fluctuations, induced by the temperature, $T$, or 
the tunneling, $t$.  This is demonstrated in the cuts shown in Fig.~\ref{fig:phase_diagram_cuts}
(indicated as dashed lines in Fig.~\ref{fig:phase_diagram}), where we also 
compare the CTQMC results with those of IPT.  The excellent agreement  
of these two approaches validates the latter, approximate method. 

The role of thermal and quantum  fluctuations is not quite identical.
In the classical ($t\to0$)  limit, $V_C^\text{class} \sim \sqrt{T \,W}$, while 
in the quantum case ($T\to0$), the escape rate $\Gamma \sim t^2/W$ takes over the role 
of the temperature, and  $V_C^\text{quant}\sim t$.

At finite temperatures,  quantum fluctuations and  thermal fluctuations 
 compete with each other. 
As  demonstrated in Fig.~\ref{fig:phase_diagram_cuts}(b),  at a finite temperatures, 
small quantum fluctuations with $\Gamma\lesssim T$ do not change the critical interaction 
strength, $V_C$,  and the transition is  mostly induced by just thermal fluctuations. 
For $\Gamma\gtrsim T$, i.e.,   $t/W\gtrsim \sqrt{T/W}$, however, quantum fluctuations play 
the dominant role, as evidenced by the almost linear shift 
of $V_C$ with increasing $t/W$.

\section{Electron glass phase: full replica symmetry breaking}\label{sec:rsb}

The exact solution of the self-consistency equations in the presence of full replica symmetry breaking is a  demanding task.  One first needs 
 to solve the non-local quantum impurity problem, Eq.~\eqref{eq:Seff} for a relatively large set of $\etilde$ values, 
extract the expectation values $\langle\delta n\rangle_{\etilde} $ as well as the dressed local 
Green's functions and susceptibilities.   Then one needs to 
solve the above-mentioned   flow equations in replica space to update the distribution $\widetilde P(\etilde)$, compute the 
average susceptibilities and Green's functions using \eqref{eq:average}, and then close the cycle
by Eq.~\eqref{eq:Seff}. 
Although this is, in principle, possible at a given point in  parameter space
using, e.g., continuous time quantum Monte-Carlo methods~\cite{WorkInProgress}, it appears to be unavoidable
to use an approximate scheme such as IPT if one aims at determining the complete phase diagram. 
Below, we summarize the results of IPT computations. Further CTQMC results shall be published 
elsewhere~\cite{WorkInProgress}. 

\subsection{Overlap distribution}\label{subsec:pq}

\begin{figure}[t!]
\includegraphics[width=0.90\columnwidth]{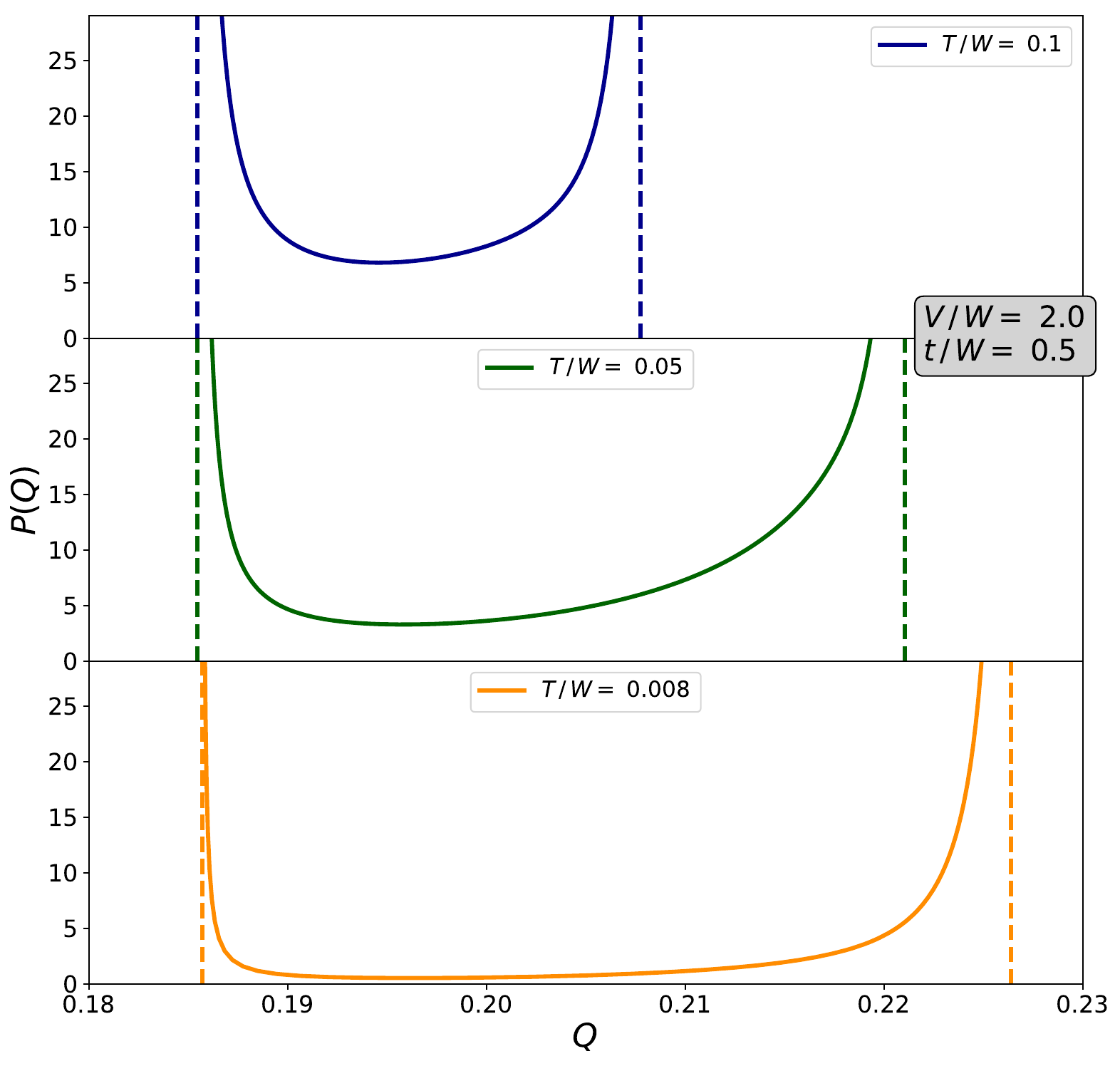}
\caption{Overlap distribution $P(Q)$ in the electron glass phase, as a function of temperature.}
\label{fig:P(Q)}
\end{figure}

The differential of the inverse function $x(Q)$ turns out to be just the distribution of the overlaps
between different replicas, 
\begin{eqnarray}
&&Q^{ab} =   \lim_{N\to\infty}  { \frac{1}{N}  \sum_i  \langle \delta n_i \rangle_a  \langle \delta n_i \rangle_b}\;,
\nonumber
\\
&& P(Q^{ab}=Q)    =  \frac{\text{d}x} {\text{d}Q }\;.
\end{eqnarray}
The numerically computed function $P(Q)$ is presented  in 
Fig.~\ref{fig:P(Q)}. In the Fermi liquid phase (not shown), $P(Q)$ is trivial, and consists of a delta function, 
$P^\text{RS}(Q) = \delta(Q-Q_\text{RS})$. This distribution indicates a unique, symmetric mean field solution. In the electron glass phase, the distribution 
becomes non-trivial, and possible overlaps have a range, $Q\in[Q_\text{min},Q_\text{max}]$, showing the onset of many symmetry-broken states.  
This overlap window becomes broader and broader as the temperature is lowered, and 
at the same time, the distribution gets depleted, and has a hight $\sim T$. This is in line with the 
observation, that at $T=0$ temperature, replica symmetry is restored. 
 (It is, however, not so clear if  a valid  expansion around this limit exists~\cite{ReadSachdevYe95}.)  Notice that the 
maximal value, $Q_\text{max}(T\to 0)$ remains less than $1/4$; this is a consequence 
of  quantum fluctuations, which tend to reduce the overlaps.

\begin{figure}[t!]
	\includegraphics[width=0.90\columnwidth]{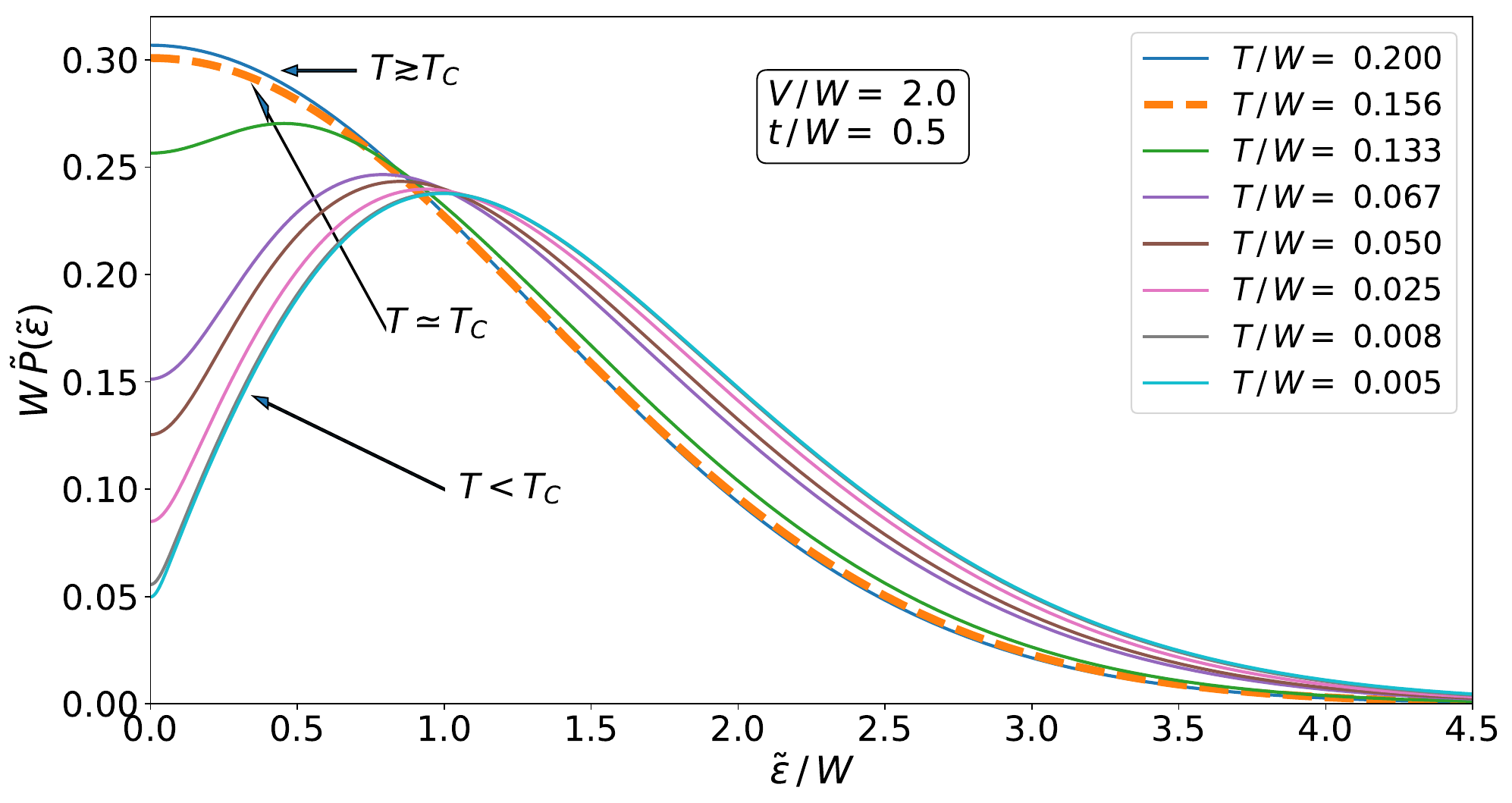}
	\caption{Formation of the pseudogap in the distribution $\Ptilde(\etilde)$ as a function of temperature.
	At $T>T_C$ the distribution remains Gaussian, but as we decrease the temperature below $T_C$ a
	pseudogap develops gradually. The thick dashed line represents the  distribution at the
	 critical temperature $T= T_C$.}
	 \label{fig:Pode}
\end{figure}

\subsection{Distribution of Hartree-Fock levels and tunneling spectra}\label{subsec:py}

As in  classical spin glasses~\cite{SK_book,Muller,Pankov2005}, a clear signature of the glass transition is the emergence of a 
Coulomb gap structure in the distribution of Hartree-Fock energies,  $\widetilde P(\etilde)$, 
shown in Fig.~\ref{fig:Pode}.  The Coulomb gap starts to open up gradually after  crossing the phase transition, and a fully developed Coulomb gap  appears only deep in the  glassy phase~\cite{Note3}. 
Although the pseudogap
 gets deeper and deeper as the temperature decreases,   $\widetilde P(\etilde=0)$ remains finite even at 
 $T=0$ temperature for any finite $t$. 
 This is a property  of  the infinite coordination limit, $z\to\infty$, where Anderson localization is absent, and 
 a  disordered metal state emerges  in the absence of interactions at $T=0$ temperature. 
Nevertheless, interactions larger than a critical  value  drive a phase transition to a replica symmetry broken 
 phase, where the density of states is strongly suppressed but finite even at $T=0$ temperature. 
 We can  interpret this phase as a metallic Coulomb glass.

\begin{figure}[t!]
	\includegraphics[width=0.90\columnwidth]{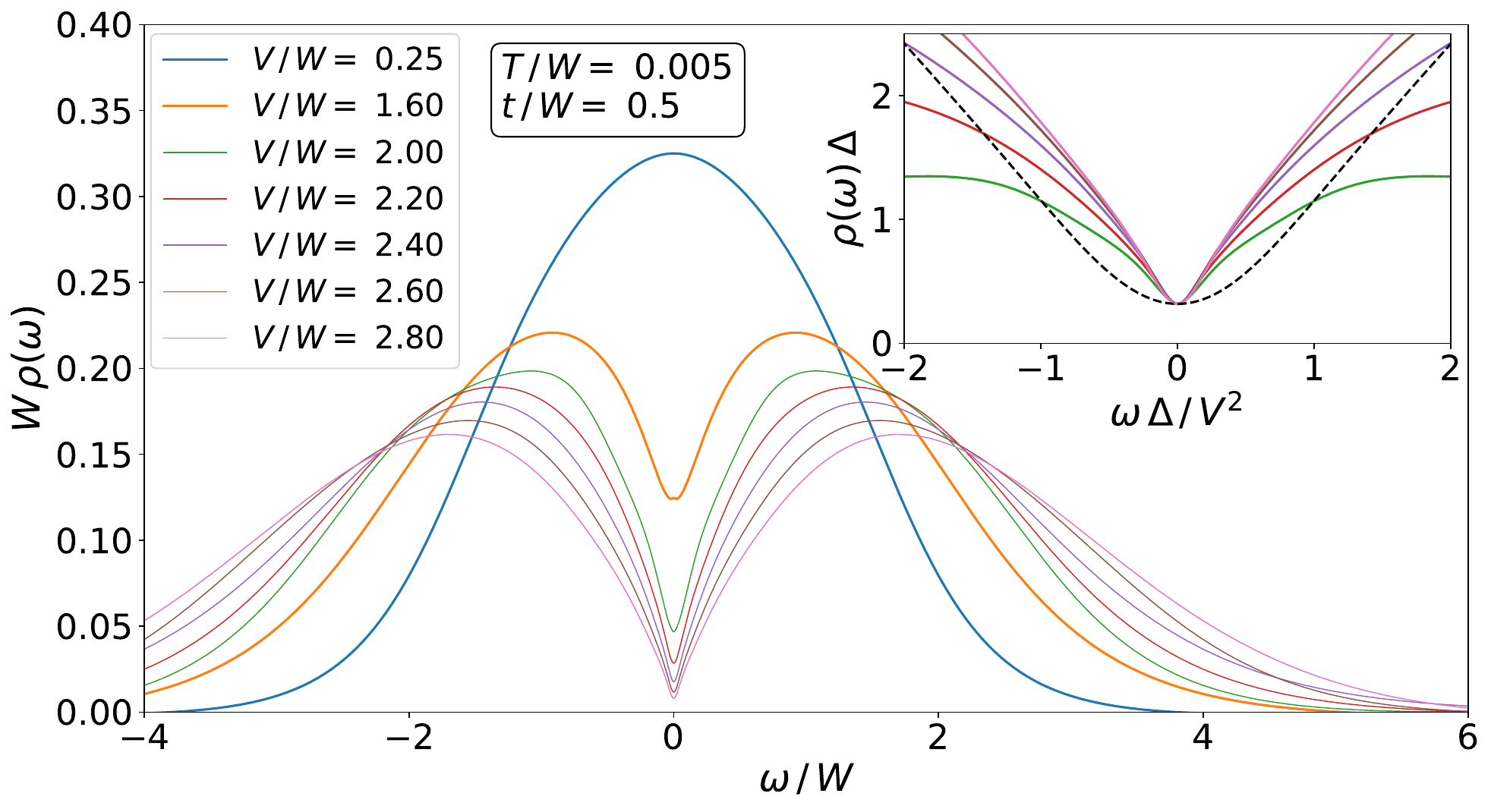}
	\caption{(Main panel) Evolution of the density of states for 
	various  interaction strengths $V$. The critical coupling is $V_c\approx 1.31\,W$.
	When $V>V_C$ a pseudogap appears in $\varrho(\omega)$, that grows with increasing V, but 
	the density of states at the Fermi level   $\varrho(0)$ always remains finite. 	
	  (Inset) Universal collapse of 
	$\varrho(\omega)$.  The dashed curve indicates the scaling curve in the classical limit, $t \to 0$.
	}
	\label{fig:density}
\end{figure}

 While the distribution of the renormalized energies  $\etilde$ is conceptually interesting,
 excepting the classical limit, their  distribution  is not  directly measurable. What is, however, 
 measurable is the tunneling density of states at a given site of renormalized energy, $\etilde$, 
 \begin{equation}
\rho_\etilde (\omega)=\dfrac{1}{\pi}\,{\rm Im}G_\etilde (\omega^+)\;,
\end{equation}
and the average density of states, $\rho(\omega)=\,{\rm Im}G(\omega^+)/\pi$, 
\begin{equation}
\rho(\omega) = \int \text{d} \etilde \; \rho_\etilde (\omega)\; \tilde P (\etilde) \, \;.
\end{equation}

Fig.~\ref{fig:density} shows the formation of the pseudogap 
in $\rho(\omega)$
at very small temperatures, as interactions are increased.
The density of states at the Fermi energy is finite, and defines a natural energy scale
$\Delta\equiv \varrho^{-1}(0)$. This scale becomes smaller and smaller 
upon increasing interactions,   while {$\rho(\omega) $} develops 
 universal scaling  as a function of  $\omega \cdot \Delta/V^2$ 
at low energies, where  it crosses over from a constant  to a linear regime, 
$\rho(\omega) \propto \omega/V^2 $ (see inset in Fig.~\ref{fig:density}).
Notice that the presence of disorder does not influence this slope, also indicating 
that the phase transition we observe is driven by interactions and not 
by disorder. The classical scaling function corresponding to $t=0$, also displayed in the inset of Fig.~\ref{fig:density}, yields the same slope as the quantum version, but the two scaling functions clearly differ, thereby demonstrating the 
difference between the role of thermal and quantum fluctuations.  
As shown in Appendix ~\ref{app:Puniv}, the distribution  $\Ptilde(\etilde)$ exhibits similar universal scaling structure.

\begin{figure}[t!]
	\includegraphics[width=0.90\columnwidth]{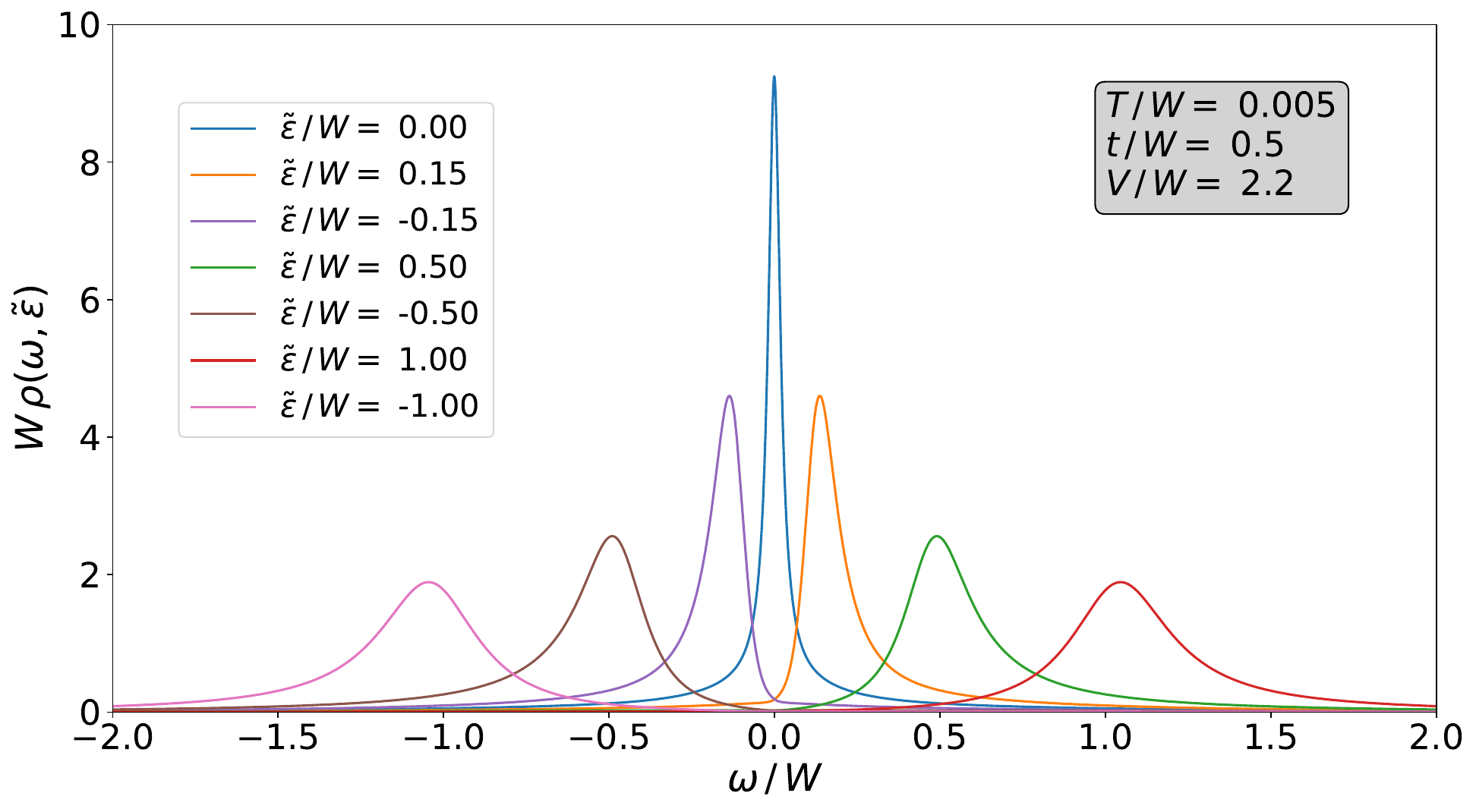}
	\caption{Unaveraged density of states deep in the glass phase for various values 
	of $\etilde$. 
	}
	\label{fig:rho_epsilon(omega)}
\end{figure}

It is instructive to investigate the structure of individual tunneling spectra, 
$\rho_\etilde (\omega)$, shown for a set of levels deep in the quantum glass regime 
in Fig.~\ref{fig:rho_epsilon(omega)}. The local density of states displays peaks at around 
the renormalized level, $\etilde$, which is broadened by quantum fluctuations. 
Levels close to the Fermi level become sharp since surrounding sites have a suppressed density of states at the 
pseudogap.

\section{Thermodynamics}\label{sec:thermo}

To determine the free energy of the glass we first need to determine  the (negative) free energy 
density $\Phi_{\rm loc}(T)$ of the effective \emph{replica action} $S_{\rm rep}$, Eq.~\eqref{eq:Srep},
\begin{equation}
\Phi_\eff(T)      \equiv \lim_{n\to 0} \frac 1 n  k_B T\, \log\,\bigg \{ \int \mathcal{D}\overline c \, \mathcal{D}c\;e^{-S_{\rm rep}[\,\overline c,c\,]}
\,\bigg\} 
\;.
\label{eq:f_rep}
\end{equation} 
This is slightly different from the  physical free energy density of the lattice model, Eq.~\eqref{eq:Ham}, which we 
denote  by $\Phi_{\rm phys}(T)$, since we must restore some terms that we threw away in course of  the 
Hubbard-Stratonovic transformation.  Restoring these  terms, which depend   on the local Green's function and 
susceptibility, we obtain
\begin{align}
\label{eq:phi_phys}
\Phi_{\rm latt}(T)& = \Phi_{\rm loc}(T) +\dfrac{t^2}{2}\int_0^\beta{\rm d}\tau\, G(\tau)G(-\tau)\nonumber\\
& -\dfrac{V^2}{4}\int_0^\beta{\rm d}\tau\, \chi(\tau)^2-\dfrac{\beta V^2}{4}\dfrac{1}{n}\sum_{b\neq a} Q_{ab}Q_{ba}.
\end{align}

In the replica symmetrical (Fermi liquid) case, Eq.\eqref{eq:f_rep} simplifies, and we obtain  
\begin{equation}
\Phi_\eff^\text{\,RS}(T) = \int \text{d}\etilde\, \Ptilde_\text{RS} (\etilde) \,\Phi_\etilde( T), 
\label{eq:phi_RS}
\end{equation}
with  the  free energy $\Phi_\etilde(T)$ computed from the local effective action,  Eq.~\eqref{eq:Seff},
and $ \Ptilde_\text{RS}$ the Gaussian Hartree level  distribution,  
displayed below Eq.~\eqref{eq:Seff}.  In this  case, the last term of 
\eqref{eq:phi_phys} also simplifies  to 
$$
-\frac 1 n \dfrac{\beta V^2}{4}\sum_{b\neq a} Q_{ab}Q_{ba} \to \beta \frac {V^2}{4} Q_\text{RS}^2\;
$$
in the $n\to 0 $ limit, yielding a complete expression for the lattice Free energy.

This procedure can be extended to the glassy phase, too, as outlined in Appendix ~\ref{app:rsb}, 
only the computation of $\Phi_{\rm loc}(T)$ becomes more complex, since one cannot decouple replicas with a single 
Hubbard-Stratonovich transformation. One still has to solve  local effective action $S_\etilde$ at the start, 
(with $\tilde \chi = \chi - Q_\text{RS}$ replaced by $ \chi - Q(1)$), and solve  
the so-called flow equations in replica space (see Appendix ~\ref{app:rsb} for details)
to obtain an expression for $\Phi_\eff(T) $  analogous to Eq.~\eqref{eq:phi_RS}. 

\begin{figure}[t!]
	\includegraphics[width=0.90\columnwidth]{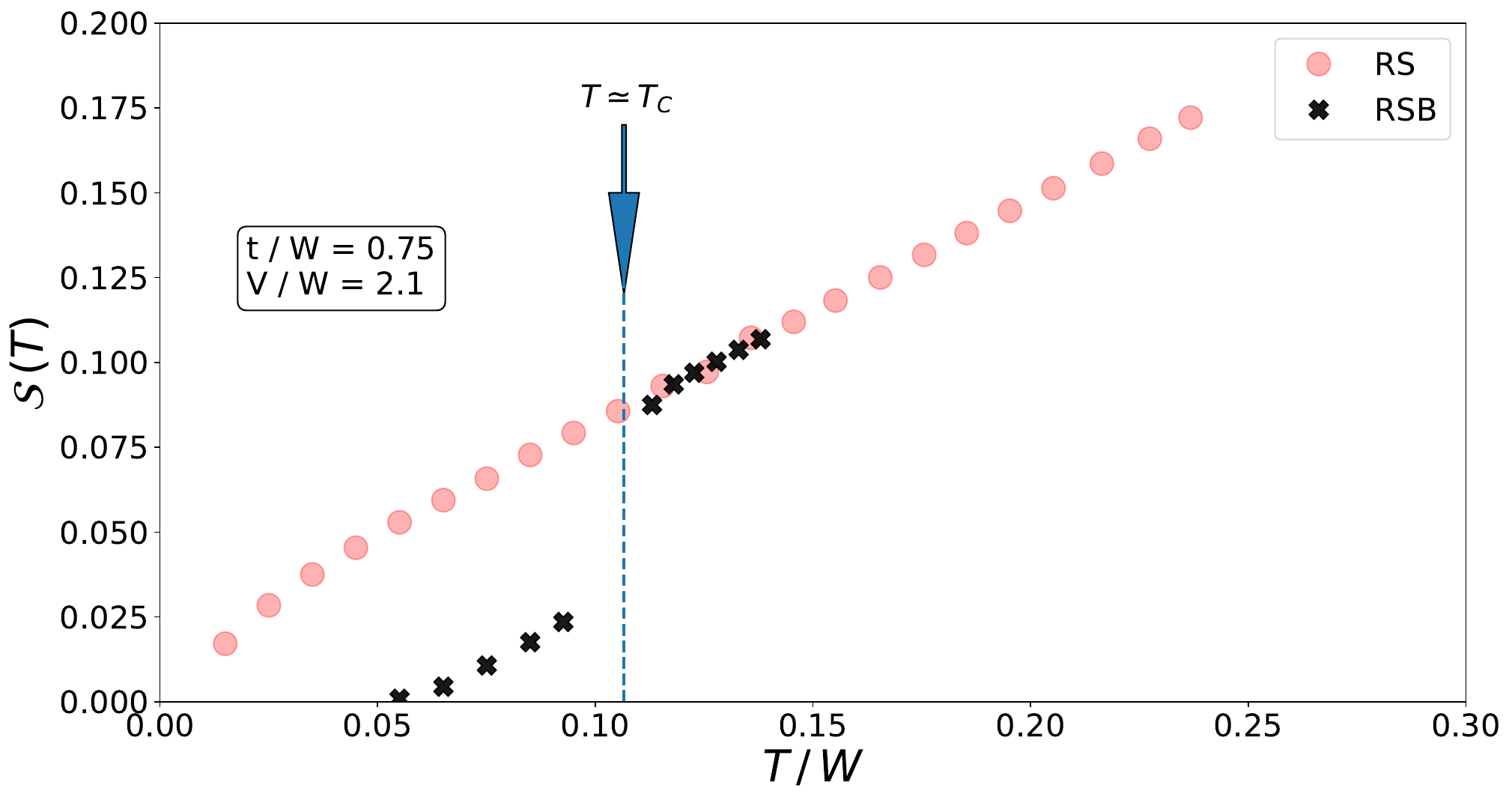}
	\caption{Temperature dependence of the entropy for the replica symmetric (RS) and replica symmetry beaking (RSB) solution.
	The RSB solution  indicates an entropy scaling to zero as $T\rightarrow 0$. Since we can not access the immediate vicinity of the critical temperature $T_C$ with the RSB calculation,  $\mathcal{S}(T)$ can not distinguish between a continuous phase transition and  a first order entropy jump at $T_C$.}
	\label{fig:entropy}
\end{figure}

Once $\Phi_\eff(T)$ and the converged Green's functions and the order parameter $Q(x)$  at hand, 
the thermodynamic quantities of the lattice model can then be computed  from
$\Phi_{\rm latt}(T)$. In particular, we determined the temperature dependent entropy density $\mathcal{S}(T)$, given by 
\begin{equation}\label{eq:entropy}
\mathcal{S}=\dfrac{\partial \Phi_\text{latt}}{\partial T}.
\end{equation}
Our results for $\mathcal{S}(T)$ are displayed the temperature dependent entropy $\mathcal{S}(T)$  in Fig.~\ref{fig:entropy}. Although we cannot decrease the temperature  
very deep down into the RSB phase, the numerical data are consistent with 
the entropy remaining positive and going quadratically to zero as $T\rightarrow 0$, corresponding to a quadratic specific heat.

Calculating the entropy $\mathcal{S}(T)$ numerically in the vicinity of the phase boundary is very challenging, due to the slow convergence of the iterative solution in the RSB phase, as well as because of the difficulties in evaluating the numerical derivative in Eq.~\eqref{eq:entropy} with high enough precision. For these reasons, we were unable to determine $\mathcal{S}(T)$ in the immediate vicinity of the critical temperature $T_C$, preventing us from answering the extremely difficult question about the order of the phase transition. The results plotted in Fig.~\ref{fig:entropy} are consistent either with a continuous phase transition or with a weakly first order transition with a latent heat. 

\begin{figure}[b!]
	\includegraphics[width=0.95\columnwidth]{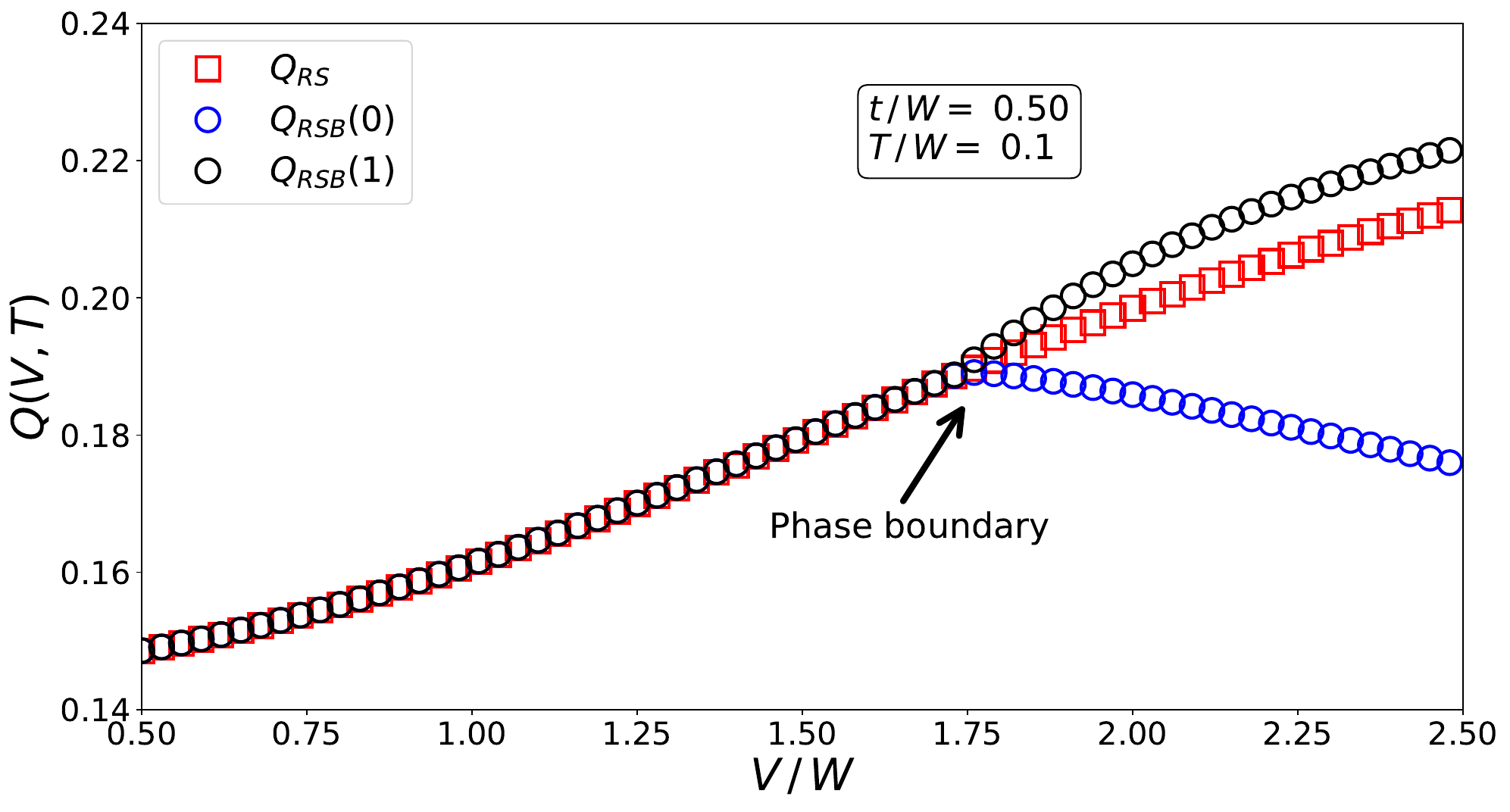}
	\caption{Overlap $Q$ across the glass transition. We plot the RS solution $Q_{RS}$, as well as the RSB results $Q_{RSB}(1)$ and $Q_{RSB}(0)$ as a function of interaction strength $V$. The glass order parameter $Q_{RSB}(1)-Q_{RSB}(0)$ remains continuous at the phase boundary, pointing towards a continuous phase transition.}
	\label{fig:Q0Q1}
\end{figure}

We can gain more insight into the properties of the glass transition by examining the behavior of the overlap function $Q(x)$ across the phase boundary. We show the RS prediction $Q_{RS}$, as well as the RSB results $Q_{RSB}(1)$ and $Q_{RSB}(0)$ as a function of interaction strength $V$ in Fig.~\ref{fig:Q0Q1}. We find that $Q_{RSB}(1)-Q_{RSB}(0)$, serving as an order parameter for the glass transition, changes continuously at the phase boundary. Similarly, the distribution $\Ptilde$ is apparently continuous through the 
Coulomb glass phase transition. We also note that we do not see any evidence for a mixed phase between the RS and RSB regimes in our numerical simulations, which could signal a first order phase transition. These findings point towards a continuous Coulomb glass transition, despite the somewhat inconclusive results in Fig.~\ref{fig:entropy}. We note that various previous studies have relied on the assumption of a continuous phase transition, by applying Landau theory to examine the glass transition~\cite{Landau1,Landau2}.

\section{Discussion}\label{sec:discuss}

We presented here a detailed study of the mean field 
Coulomb glass (disordered $t-V$)  model of Ref.~\cite{Pastor1999} in the quantum regime, 
in the Fermi liquid (replica symmetrical)  as well as deep 
in the glassy (replica symmetry) broken phase.
The combination of  continuous time quantum Monte-Carlo approach with
  Iterative Perturbation Theory (IPT)  allowed us to map accurately the phase 
  boundaries separating the interaction induced glassy phase from the Fermi liquid phase 
  in the classical as well as in the quantum regime, and to determine spectral 
  functions as well as thermodynamic properties. 
 Having validated our IPT scheme in the metallic regime, we used it to 
enter the electron glass phase, where complete replica symmetry breaking 
must be incorporated in the theory. 

In the spectral function, we observe the formation of a plasmonic correlation hole 
in the average tunneling density (local density of states) already in the Fermi liquid phase. 
This correlation hole smoothly develops into an   Efros-Shklovskii pseudogap when we enter 
the electron glass phase, where replica symmetry is broken.  The Efros-Shklovskii pseudogap gap is, however, not 
fully developed in this mean field model even at $T=0$ temperature: 
similar to thermal fluctuations, small quantum fluctuations induce a finite   
density of states even at the Fermi energy. This  is  a peculiarity of 
the infinite coordination limit, where Anderson  localization is absent, and a glassy 
Fermi liquid state emerges rather than a glassy localized phase. 
For small tunneling, the average density of states exhibits 
universal scaling  at at low temperatures and low energies. 
Similar features have been observed in the transverse field Sherrington-Kirkpatrick model 
in Ref.~\cite{Andreanov2012}.

We have also computed the local density of states. In the electron glass phase, this 
typically consists  of sharp resonances, located around some renormalized 
Hartree energies, whose distribution also exhibits a pseudogap. 
These resonances become  sharper and sharper as one 
approaches the Fermi energy, but retain their finite width, even at the 
Fermi surface, indicating again 
that these states remain extended even in the glass phase. 

We have also constructed the full thermodynamic mean field description of the disordered $t-V$ model, and 
analyzed its thermodynamic properties in both phases. 
We obtain a continuous free energy, however, our present results are   not  yet accurate enough to distinguish between a continuous phase transition and a first order entropy jump. While the entropy calculation remains inconclusive, the apparent continuous behavior of the glass order parameter at the phase boundary, as well as the lack of numerical evidence for a mixed phase characteristic of first order transitions, points towards a continuous glass transition, in accordance with  the results of a Landau 
functional approach~\cite{Landau1,Landau2}. 
 Further, very accurate continuous time quantum Monte-Carlo 
computations for the entropy within the replica symmetry broken phase could provide more evidence for this conclusion.

As mentioned above, the absence of the insulating electron glass phase is 
an artifact of the infinite coordination limit. However, the unavoidably 
metallic  glass phase emerging in this model  is relevant for many 
 metallic  disordered systems, which exhibit glassy 
behavior. Amorphous polycrystaline solids~\cite{Ovadyahu2007} or 
 granular metals~\cite{Delahaye2008} are such examples, 
but  metallic electron glass phases  can  be observed in certain 
two dimensional systems~\cite{Popovic2002,Popovic2002b}.
Thorough studies 
of $\rm{Na}^{+}$ doped silicon MOSFETs 
reveal  a metal insulator transition at a carrier density, $n\approx n_c$, and 
an intermediate  metallic  glass phase emerges on the metallic side of the transition
at concentrations $n_g>n> n_c$,  as evidenced 
by low frequency resistance noise~\cite{Popovic2002b} and ageing~\cite{DraganaPopovic_memory} experiments
on low mobility samples. A metallic glass phase could also be experimentally realized in ultracold atomic settings, by placing fermionic atoms into a multi-mode cavity ~\cite{Muller2012LongRange}.

The understanding and solution of the mean field Coulomb glass model, Eq.~\eqref{eq:Ham},
is just a first step in constructing a mean field theory of the real Coulomb glass. 
In fact, it is quite unclear, how one could incorporate  localization 
and long-ranged interactions at the same time in a mean field model.
To have an Anderson-localized phase, one should  impose 
a finite coordination number, $z$, and thereby exclude the presence of infinite 
number of nearest neighbors. This challenging problem can be studied via an extended dynamical mean field theory (EDMFT) approach~\cite{EDMFT_PhysRevB.66.045117} and a coherent potential approximation (CPA), allowing to
use the present scheme as a \emph{local approximation} to describe the glassy 
phase transition in a finite dimensional system~\cite{Muller}.  Alternatively, Anderson localization can also be captured by combining DMFT arguments with a typical medium theory (TMT) scheme~\cite{TMT_Dobrosavljevi__2003}.
 
 Another open question is that of glassy dynamics. 
Global charge response should reflect  the emergence of a glassy phase through anomalously slow response and a 
broad distribution of scales~\cite{OregImry_review,Pollak2002}. It remains an open question, 
how the present approach is able to explain this behavior.  Finally, spin degrees of freedom 
have  been completely neglected in this work.  The role of  Mott-Anderson 
physics  and spontaneous spin formation should be further explored and elucidated. 

Although in this work we only focused on the description of the Coulomb glass phase,
the method and formalism presented pave the way to study quantum correlations 
in the glassy phase of many  mean field quantum glass models, 
such as the transverse field Sherrington-Kirkpatrick model, and the disordered 
Dicke model. 
All these questions are and should be subject of future research.

\begin{acknowledgments}
We thank Vladimir Dobrosavljevi\'c and Pascal Simon  for  insightful  discussions. This work has been supported by the National Research Development and Innovation Office (NKFIH) through Grant Nos. SNN118028 and K124176, within the Quantum Information National Laboratory, and by the European Research Council (ERC) under the European Unions Horizon 2020 research and innovation program (Grant Agreement No. 771537). CPM was supported by UEFISCDI under project No. PN-III-P4-ID-PCE-2020-0277, and the project for funding the excellence, contract No. 29 PFE/30.12.2021.
\end{acknowledgments}

\appendix

\section{Details of continuous time quantum Monte Carlo method}\label{app:CTQMC}


In this appendix we present the derivation of the combined Monte Carlo weight $w(z_{k}) = {\rm det}\hat{F}^{(k)}\,w_{\tilde{\varepsilon}}\,w_{\tilde{\chi}}$ based on the implementation presented in Ref.~[\onlinecite{werner-EDMFT}].

We separate the effective local action given in Eq.~(\ref{eq:Seff}) as
\begin{eqnarray}
S_{\etilde} &\equiv& S_{\rm F} + S_{1},\\
S_{\rm F} &=& -\int_\tau \int_{\tau^\prime} 
\overline{c}_\tau t^2 \; G(\tau-\tau^\prime) \; c_{\tau^\prime}, \label{SF}\\
S_{1} &=& \int_\tau \overline{c}_\tau ( \partial_{\tau} +\tilde{\varepsilon} ) c_\tau \nonumber\\
&-& \frac{V^2}{2} \int_\tau \int_{\tau^\prime}\large( \chi(\tau-\tau^\prime)- Q_\text{RS}\large)\; \delta n_\tau \;\delta n_{\tau^\prime} ,
\label{S1}
\end{eqnarray}
and expand the partition function $Z_\etilde={\rm Tr}\,{\rm e}^{-S_{\etilde}}$ in terms of the hybridiztion part  $S_{\rm F}$, which gives
\begin{eqnarray}
Z_\etilde &=& {\rm Tr}\,{\rm e}^{-(S_{\rm F}+S_{1})} =\\
&=& \sum_{k}  \int_{0}^{\beta}dt\tau_{1}...d\tau_{k} \int_{0}^{\beta}dt\tau^{\prime}_{1}...d\tau^{\prime}_{k}\, {\rm det} \hat{F}^{(k)}
\nonumber\\
&\times& {\rm Tr}\left[\,{\rm e}^{-S_{1}}\, \overline{c}_{\tau_{1}} c_{\tau^{\prime}_{1}}  ... \overline{c}_{\tau_{k}} c_{\tau^{\prime}_{k}}\right] = \int {\cal D}(k)w(z_{k}),
\end{eqnarray}
where we introduced the notation $ \int {\cal D} = \sum_{k}  \int_{0}^{\beta}dt\tau_{1}...d\tau_{k} \int_{0}^{\beta}dt\tau^{\prime}_{1}...d\tau^{\prime}_{k}$, and therefore the weight $w(z_{k})$ is expressed as
\begin{eqnarray}
w(z_{k}) &=& {\rm det} \hat{F}^{(k)}\,  {\rm Tr}\,\left[{\rm e}^{-S_{1}} \,\overline{c}_{\tau_{1}} c_{\tau^{\prime}_{1}}  ... \overline{c}_{\tau_{k}} c_{\tau^{\prime}_{k}}\right] \nonumber\\
 &=& {\rm det} \hat{F}^{(k)}\,  \langle \overline{c}_{\tau_{1}} c_{\tau^{\prime}_{1}}  ... \overline{c}_{\tau_{k}} c_{\tau^{\prime}_{k}} \rangle.
\end{eqnarray} 
By evaluating the first term of $S_{1}$ in Eq.~(\ref{S1}) in the segment picture, we obtain the weight $w_{\tilde{\varepsilon}}$  as 
\begin{eqnarray}
w_{\tilde{\varepsilon}} = {\rm e}^{-\tilde{\varepsilon}\, l},\label{wtildeepsilon}
\end{eqnarray}
where $l=\sum_{i=1}^{k}l_{i}$ is the total length of the segments $l_{i}$ with $l_{i} =\tau_{i} -\tau^{\prime}_{i} $.

The weight contribution from the second term in Eq.~(\ref{S1}) is expressed as
\begin{eqnarray}
w_{\tilde{\chi}} &=& {\rm e}^{ \frac{1}{2}V^2 \int_{0}^{\beta} d\tau_{1} \int_{0}^{\beta} d\tau_{2} \tilde{\chi}(\tau_{1}-\tau_{2})\, \delta n_{\tau_{1}} \delta n_{{\tau^\prime_{2}}}} \label{wtildechi}.
\end{eqnarray}
Defining a function $K(\tau)$ as $K(\tau)^{\prime\prime} = V^2 \tilde{\chi}(\tau)$ with the conditions $K(0)=0$ and $K^{\prime}(0)=1/2 V^2\int_{0}^{\beta} d \tau \tilde{\chi}(\tau)$, the integral in Eq.~(\ref{wtildechi}) is evaluated as
\begin{eqnarray}
w_{\tilde{\chi}} &=& 
{\rm exp} \biggl( \frac{1}{2} \sum_{k_{1}, k_{2}}  \left[ -K(\tau_{k_1}^{\prime}-\tau_{k_2}^{\prime}) + K(\tau_{k_1}-\tau_{k_2}^{\prime})
\right. \biggl. \nonumber \\ 
  &+&  \biggl. \left.  K(\tau_{k_1}^{\prime}-\tau_{k_2})  - K(\tau_{k_1}-\tau_{k_2}) \right]
 \biggr. \nonumber\\ 
  &+& \biggr. K^{\prime}(0)l (1 - 2\langle n \rangle) \biggr),
\label{eq-wchi2}
\end{eqnarray}
which can be rewritten as 
\begin{eqnarray}
w_{\tilde{\chi}} &=& {\rm exp} \biggl( -\sum_{i>j} s_{i}s_{j} [ K(\tilde{\tau}_{i}-\tilde{\tau}_{j}) - K(0) ] 
\biggr. \nonumber\\
&+& \biggl. K^{\prime}(0)l (1 - 2\langle n \rangle) \biggr) \nonumber\\
&=& {\rm exp} \left( -\sum_{i>j} s_{i}s_{j}  K(\tilde{\tau}_{i}-\tilde{\tau}_{j}) \right),
\label{eq-wchin}
\end{eqnarray}
where the times are ordered as $0<\tilde{\tau}_{1}<\tilde{\tau}_{2}< ... < \beta$, and $s$ is $+1$ for a creation operator and $-1$ for  annihilation operator.
Thus, $w(z_{k})$ can finally be expressed in the compact form $w(z_{k}) = {\rm det} \hat{F}^{(k)}\,w_{\tilde{\varepsilon}}\,w_{\tilde{\chi}}$.
We note that quantities can be measured without additional computation cost compared to the case of $V=0$ with this modified weight.

The Green's function $G(\tau) = \langle T_{\tau}\, c(0)\, \overline{c}(\tau) \rangle$ is evaluated in the same way in our Monte Carlo procedure as in the absence of the retarded interaction $V^2\tilde{\chi}(\tau-\tau^{\prime})$. 
Namely, for measuring the Green's function we need a configuration where operators $c_{\tau^\prime}$ and $\overline{c}_{\tau}$ are unconnected. In the hybridization method this is achieved by removing one of the hybridization
lines, resulting in
\begin{eqnarray}
G(\tau) = \left\langle \frac{1}{\beta} \sum_{i,j}^{k} \left( \hat{F}^{(k)} \right)^{-1}_{ji} \tilde\delta(\tau, \tau_{i}-\tau^{\prime}_{j})  \right\rangle_{\rm MC}.
\end{eqnarray}
Here $F_{i,j}^{(k)} = F(\tau_{i}-\tau^{\prime}_{j})$ is the hybridization matrix, and we have defined
\begin{equation*}
\tilde{\delta}(\tau,\tau^\prime)=
\begin{cases}
\delta(\tau-\tau^\prime)&{\rm if }\quad\tau^\prime>0\\
-\delta(\tau-\tau^\prime)&{\rm if}\quad\tau^\prime<0.
\end{cases}
\end{equation*}

Both averaged susceptibilities 
$\chi(\tau) = \langle \delta n_{\tau} \delta n_{0} \rangle$ and
$\tilde{\chi}(\tau) = \langle n_{\tau} n_{0}  \rangle - \langle n \rangle^2$
can be sampled in the Monte Carlo simulation, and therefore the properties  
$\chi(\tau=0)=1/4$, $\tilde{\chi}(\tau=0) = 1/4 - q_{\rm RS}$, and $\tilde{\chi}(\tau) = \chi(\tau) - q_{\rm RS}$ can be check-points for the correctness of the CTQMC code.

Our choice for the zeroth order ansatz for $G^{[0]}(\tau)$, $\tilde{\chi}^{[0]}(\tau)$, and $Q^{[0]}_{\rm RS}$ 
in obtaining the self-consistent replica symmetric solution are the non-interacting ones as
\begin{eqnarray}
G^{[0]}(i \omega_{n}) &=& i \frac{1}{2t^2}\left( \omega_{n} - \sqrt{4t^2 + \omega_{n}}  \right) \nonumber\\
&& \xrightarrow[]{\text{FFT}}  \,  G^{[0]}(\tau),
\\
\tilde{\chi}^{[0]}(\tau) &=&  G^{[0]}(\tau) G^{[0]}(-\tau), \\
Q^{[0]}_{\rm RS} &=& 0.
\end{eqnarray}
We note that the choice of the ansatz does not affect the converged result.

\section{Replica symmetry breaking}\label{app:rsb}

In the glassy phase, replica symmetry is broken, and $Q^{ab}$ acquires a non-trivial structure in replica space. 
In the  limit $n\to0$, we characterize the matrix $Q^{ab}$ in terms of a continuous variable, 
$x\in [0,1]$, and a corresponding function,   $Q_{ab}\rightarrow Q(x)$. 
The parameter $x$ in this language 
characterizes  deeper and deeper levels  of replica symmetry breaking as $x$ flows from 0 towards 1.

As stated in the main text, the simple construction leading to Eqs.~\eqref{eq:Seff} and ~\eqref{eq:average} can be 
generalized  to this more complicated case, too.  Following steps similar  to those in Refs.~\cite{Sommers_1984,Muller}, 
we can introduce a set of effective one level  models (actions), parametrized by $x$, and describing different levels
of replica symmetry breaking,
 restricted   free energy densities,    $\phi_{\etilde,x}(T)$, and corresponding level distributions,  $P_x(\etilde)$, 
 both temperature dependent quantities.
 
 There is a  trade-off between these two quantities: at $x=1$, $\phi_{\etilde,x}$  simplifies 
 to 
 \begin{equation}
\phi_{\etilde,x=1}   = \Phi_\etilde(T) =\dfrac{1}{\beta}\log Z_{\etilde},\nonumber\\
 \end{equation}
where $Z_{\etilde}$ is computed from the effective action, Eq.~\eqref{eq:Seff}, 
with  
\begin{equation}\label{eq:chitilde}
\tilde\chi(\tau-\tau^\prime)\equiv \chi(\tau-\tau^\prime)-Q(1). 
\end{equation}
At the same time,  the distribution $\Ptilde_{x\to 1}(\etilde)$ has a complicated, renormalized  form
 \begin{equation}
\Ptilde_{x\to 1}(\etilde)  \equiv \Ptilde(\etilde),
 \end{equation}
i.e., the distribution, which enters the computation of the average Green's function.
 
In contrast, for $x=0$, the distribution $\Ptilde_{x\to 0}(\etilde)$ becomes  just the bare 
distribution of levels (without replica symmetry breaking),   with $Q_\text{RS}$ replaced by $Q_0 \equiv Q(0)$,  
\begin{equation}
\Ptilde_{x\to 0}(\etilde)\equiv P_0(\etilde) 
=\dfrac{\exp\left\lbrace-\etilde^2/\left(2\left[W^2+V^2 Q_0\right]\right)\right\rbrace}{\sqrt{2\pi\left[W^2+V^2 Q_0\right]}}\;, 
\end{equation}
 while 
$\phi_{\etilde,x=0}$ incorporates all scales of replica symmetry breaking
in the range $x\in [0,1]$,  and is directly related to the physical {(negative) free energy density of the local replica action $S_{\rm rep}$, $\Phi_\eff(T)$}, as
\begin{equation}\label{eq:phi_rep}
\Phi_\eff(T) = \int \text{d} \etilde \; P_0(\etilde) \,\phi_{\etilde,x=0}\,.
 \end{equation}

The distributions $P_{\etilde,x}$ and the free energies $\phi_{\etilde,x}$ 
at different layers  of replica symmetry breaking are related
by  flow equations, which we can derive following the lines of Refs.~\cite{Muller,Pankov2005,Sommers_1984} . This 
relation is expressed in terms of simple partial differential equations: 
\begin{align}
&\partial_x
\phi_{\etilde,x}=-\dfrac{V^2}{2}\dfrac{dQ}{dx}\left\lbrace \partial_\etilde^{\,2}\,\phi_{\etilde,x}+\beta x\left(\partial_\etilde\phi_{\etilde,x}\right)^2\right\rbrace, 
\label{eq:flow_phi}
\\
&\partial_x P=\dfrac{V^2}{2}\dfrac{dQ}{dx}\left\lbrace \partial_\etilde^{\,2} P-2\beta x\,\partial_\etilde\left(P\partial_\etilde\phi_{\etilde,x}\right)\right\rbrace.
\label{eq:flow_P}
\end{align}
These equations just express the fact that one can determine $\phi_{\etilde,x}$ and $P_x(\etilde)$ at a "deeper" RSB level, 
$x-\text{d}x$,  from  the knowledge of the energy dependent  free energy at level $x$ and the corresponding distribution, 
$P(x,\etilde)$.  

Finally, {$Q(x)$ is determined from the last equation of the self-consistency condition Eq.~\eqref{eq:consistency2}, coinciding with the marginality condition, which ensures that the free energy is marginal with respect to 
all variations of  $Q(x)$.} This leads to the self-consistency equation, 
\begin{equation}
Q(x)=\int\,d\etilde P_\etilde(x)\left[\partial_\etilde\phi_{\etilde,x}\right]^2.
\label{eq:margin}
\end{equation}

Using IPT, the solution thus proceeds as follows: Having some  ansatz for $G(\tau)$, $\widetilde \chi(\tau)$, 
and $Q(x)$,  we first solve the action $S_\etilde$ in Eq.~\eqref{eq:Seff} {within the Hartree-Fock approximation}, 
and determine $\Phi_{\etilde,x=1}=\Phi_\etilde$   for a dense set of levels $\etilde$'s. 
We then solve Eq.~\eqref{eq:flow_phi} backwards, from $x=1$ to $x=0$ to obtain an estimate for $\Phi_{\etilde,x}$. 
Using $\Phi_{\etilde,x}$, we can now solve  Eq.~\eqref{eq:flow_P} to obtain the distributions 
$P_x(\etilde)$ from $P_{x=0}(\etilde)$. 
We then use $P_(\etilde)$  together with $\Phi_{\etilde,x}$ to estimate $Q(x)$ by the marginality condition,  \eqref{eq:margin}. 
Finally, having our estimate for   $P_{x=1}(\etilde) = \Ptilde(\etilde)$ and for $Q(1)$, 
we can use 
Eq.~\eqref{eq:average} to obtain better estimates for $\widetilde \chi(\tau)$
and $G(\tau)$. This procedure is iterated until convergence is reached. 

The most demanding part of this iteration procedure is the solution of the quantum impurity problem
for {roughly  thousand} values of $\etilde$ in each iteration step.


\section{Universal scaling of $\Ptilde(\etilde)$}\label{app:Puniv}

\begin{figure}[b!]
	\includegraphics[width=0.950\columnwidth]{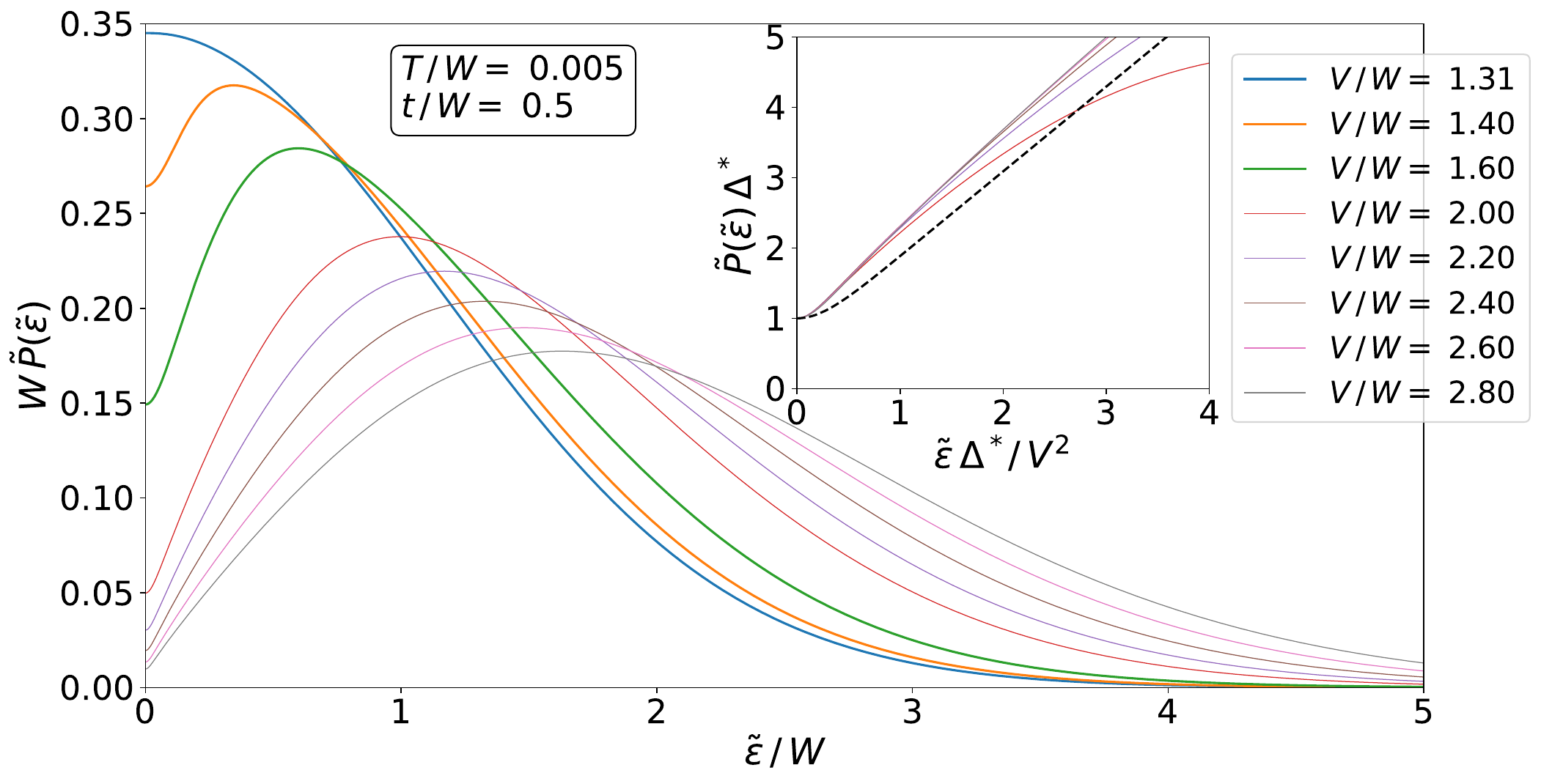}
	\caption{(Main panel) Evolution of the local field distribution in the RSB phase. The top curve
	corresponds to the critical coupling, $V_c = 1.31 W$. With increasing the interaction $V$, 
a pseudogap is formed that deepens with a slope independent of the strength of disorder $W$. (Inset) Scaling of the field distribution deep in the glassy phase.
	\label{fig:P_eps}}
\end{figure}

In the main text, we have shown that the spectral function $\rho(\omega)$ displays universal scaling in the strong interaction limit. The distribution $\Ptilde(\etilde)$ displays similar scaling properties. 
Similar to the SK model,   $\Ptilde(\etilde)$
scales linearly over an extended region  in the limit of small quantum-tunneling and temperatures, 
 $\Ptilde(\etilde)\approx 1.13\;  |\etilde|/V^2$,  with a slope independent of  the strength of disorder. 
 In the quantum limit,  $\Ptilde(0)$ remains finite even as $T\to 0$,    $\Ptilde(0)=1/\tilde \Delta$.
 Similar to $\rho(\omega)$, as shown in the inset of Fig.~\ref{fig:P_eps}, 
 $\tilde \Delta \Ptilde(\etilde)$ becomes a universal function of $\etilde \tilde \Delta/V^2$ in this quantum limit.
 Notice, however, that there seems to be no simple relation between the scales $\tilde \Delta$ and $\Delta$.

\end{document}